\documentclass[%
  aip,
 reprint,%
]{revtex4-2}

\usepackage[version=4]{mhchem}
\usepackage{graphicx}
\usepackage{dcolumn}
\usepackage{bm}

\usepackage{xcolor}
\usepackage{float}
\usepackage{amsmath}
\usepackage{amsfonts}
\usepackage{amssymb}
\usepackage[utf8]{inputenc}
\usepackage[T1]{fontenc}
\usepackage{mathptmx}
\usepackage{booktabs}

\usepackage{textcomp}

\begin{document}
\preprint{AIP/123-QED}

\title{Relativistic ultrafast electron diffraction at high repetition rates}

\author{K. M. Siddiqui}
\email{khalid.siddiqui@phys.au.dk}\altaffiliation[Present Address:~]{Department of Physics and Astronomy, Aarhus University, Aarhus, Denmark.}
\affiliation{Materials Sciences Division, Lawrence Berkeley National Laboratory, Berkeley, California 94720, USA.}

\author{D. B. Durham}
\affiliation{National Center for Electron Microscopy, Molecular Foundry, Lawrence Berkeley National Laboratory, Berkeley,
California 94720, USA.}

\affiliation{Department of Materials Science and Engineering, University of California at Berkeley, Berkeley, California 94720,
USA.}

\author{F. Cropp}
\affiliation{Accelerator Technology and Applied Physics Division, Lawrence Berkeley National Laboratory, Berkeley, California 94720, USA.}

\affiliation{Department of Physics and Astronomy, University of California Los Angeles, Los Angeles, California 90095, USA.}

\author{F. Ji}
\affiliation{Accelerator Technology and Applied Physics Division, Lawrence Berkeley National Laboratory, Berkeley, California 94720, USA.}

\author{S. Paiagua}
\affiliation{Lawrence Berkeley National Laboratory, One Cyclotron Road, Berkeley, California 94720, USA}

\author{C. Ophus}
\affiliation{National Center for Electron Microscopy, Molecular Foundry, Lawrence Berkeley National Laboratory, Berkeley,
California 94720, USA.}

\author{N. C. Andresen}
\affiliation{Engineering Division, Lawrence Berkeley National Laboratory, Berkeley, California 94720, USA.}

\author{L. Jin}
\affiliation{Department of Materials Science and Engineering, University of California at Berkeley, Berkeley, California 94720, USA.}

\author{J. Wu}
\affiliation{Department of Materials Science and Engineering, University of California at Berkeley, Berkeley, California 94720, USA.}

\author{S. Wang}
\affiliation{Department of Electrical Engineering and Computer Sciences, University of California at Berkeley, Berkeley, California 94720, USA.}

\author{X. Zhang}
\affiliation{Department of Mechanical Engineering, University of California at Berkeley, Berkeley, California 94720, USA.}

\author{W. You}
\affiliation{Department of Physics and JILA, University of Colorado and NIST, Boulder, Colorado 80309, USA.}

\author{M. Murnane}
\affiliation{Department of Physics and JILA, University of Colorado and NIST, Boulder, Colorado 80309, USA.}
 
\author{M. Centurion}
\affiliation{Department of Physics and Astronomy, University of Nebraska-Lincoln, Lincoln, Nebraska 68588, USA.}

\author{X. Wang}
\affiliation{Department of Physics and Astronomy, University of Nebraska-Lincoln, Lincoln, Nebraska 68588, USA.}

\author{D. S. Slaughter}
\affiliation{Chemical Sciences Division, Lawrence Berkeley National Laboratory, 1 Cyclotron Rd, Berkeley, California 94720, USA}

\author{R. A. Kaindl}
\affiliation{Materials Sciences Division, Lawrence Berkeley National Laboratory, Berkeley, California 94720, USA.}

\affiliation{Department of Physics, Arizona State University, Tempe, Arizona 85287, USA.}

\author{P. Musumeci}
\affiliation{Department of Physics and Astronomy, University of California Los Angeles, Los Angeles, California 90095, USA.}

\author{A. M. Minor}
\affiliation{National Center for Electron Microscopy, Molecular Foundry, Lawrence Berkeley National Laboratory, Berkeley,
California 94720, USA.}
\affiliation{Department of Materials Science and Engineering, University of California at Berkeley, Berkeley, California 94720,
USA.}

\author{D. Filippetto}
\email{dfilippetto@lbl.gov}
\affiliation{Accelerator Technology and Applied Physics Division, Lawrence Berkeley National Laboratory, Berkeley, California 94720, USA.}

\date{\today}

\begin{abstract}
The ability to resolve the dynamics of matter on its native temporal and spatial scales constitutes a key challenge and convergent theme across chemistry, biology, and materials science. The last couple of decades have witnessed ultrafast electron diffraction (UED) emerge as one of the forefront techniques with the sensitivity to resolve atomic motions. Increasingly sophisticated UED instruments are being developed that are aimed at increasing the beam brightness in order to observe structural signatures, but so far they have been limited to low average current beams. Here we present the technical design and capabilities of the HiRES (High Repetition Rate Electron Scattering) instrument, which blends relativistic electrons and high repetition rates to achieve orders of magnitude improvement in average beam current compared to the existing state-of-the-art UED instruments. The setup utilizes a novel electron source to deliver femtosecond duration electron pulses at up to MHz repetition rates for UED experiments. We provide example cases of diffraction measurements on solid-state and gas-phase samples, including both micro- and nanodiffraction modes, which showcase the potential of the instrument for novel UED experiments.

\end{abstract}

\maketitle

\begin{quotation}

\end{quotation}

\section{Introduction}
\label{section:intro}
Development of pulsed electron sources has revolutionized the field of structural dynamics leading to a new paradigm in the investigation of ultrafast phenomena. To this end, diffraction techniques have taken center stage for providing atomic-level structural information in gas and condensed phase systems. The integration of femtosecond laser technology with electrons bunches generated via photoemission from metallic cathodes was pioneered by Mourou and Williamson \cite{Mourou1982} following seminal work of Ischenko \textit{et al} \cite{Ischenko1983}, culminating in the first ultrafast electron diffraction (UED) experiment \cite{filippetto2022rmp}. Since then, ultrafast electron probes have been employed in numerous studies encompassing different materials and research themes over a broad range of time scales. Examples include discoveries of light-induced non-equilibrium phases in quantum materials \cite{Kogar2020,Gao2013}, visualization of gas-phase photochemical metamorphosis \cite{Dudek2001,Cao1999}, measurements of non-thermal melting \cite{Siwick2003,Wu2022}, warm dense matter formation \cite{Mo2016} and plasma waves \cite{Sun2020,Zandi2020}, mapping of structural dynamics in molecular crystals \cite{Jean-Ruel2013,Gao2013} and much more. 

\begin{figure*}[t!]
    \centering
    \includegraphics[width=16cm]{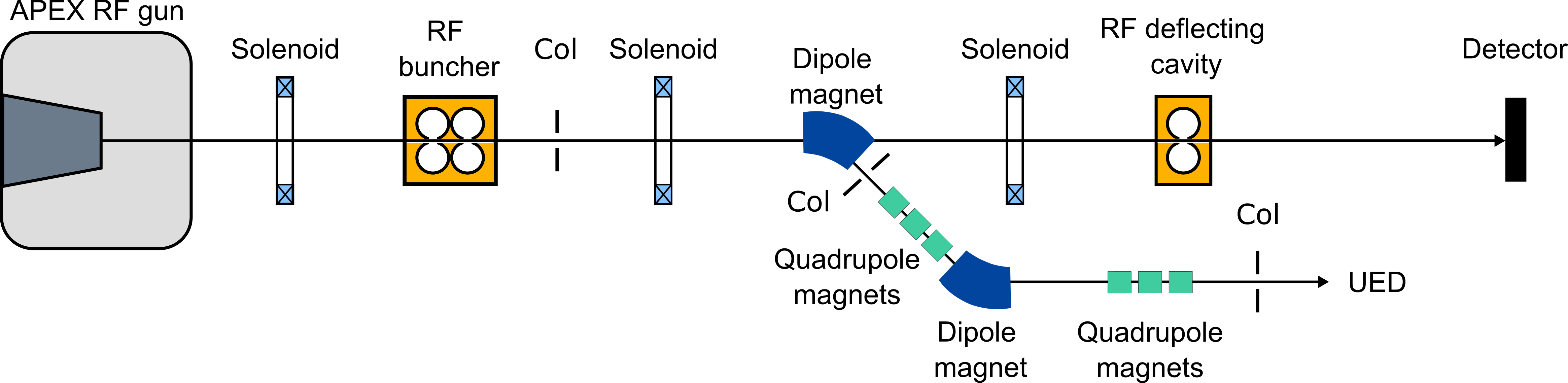}
    \caption{Layout of HiRES beamline which splits out at the dogleg for UED and diagnostic branches. Col: Collimator. }
    \label{HiRES-Beamline}
\end{figure*}

A well-known factor inherent to UED is the space charge effect due to mutual repulsion between electrons. It restricts the particle number density that can be delivered in each bunch, leads to pulse lengthening, and degrades the overall beam brightness. Several strides in theory \cite{Michalik2006,Reed2006} and instrumentation have been made in the last two decades aimed at mitigating the influence of space-charge effects. Namely, a plethora of gun designs have been developed over the years, employing increasingly advanced technologies from electron microscopes and accelerators. The first guns used for femtosecond electron diffraction were compact table-top DC guns operating in the tens of keV regime, in which the short path for the electron probe permitted use of low-charge beams (< 10000 electrons per pulse) with sub-picosecond temporal resolution \cite{Siwick2003,Storeck2021}. Modified transmission electron microscopes for producing high-coherence, single-electron pulses at high repetition rates (ultrafast electron microscopes \cite{Barwick2008,Aidelsburger2010}) or single-shot high-charge pulses (dynamic transmission electron microscopy \cite{LaGrange2006}) followed, as well as UED setups utilizing radio frequency (RF) \cite{Chatelain2012} and terahertz (THz) \cite{Zhang2021} drives for rapid acceleration and compression to minimize space-charge interactions during electron transport. More recently, the use of relativistic electron beams in UED \cite{Musumeci2008,SLAC_first} has grown as the marriage of large acceleration fields (> 10 MV/m) and high final kinetic energies can provide very high-density electron bunches with minimal pulse lengthening. 

However, despite these developments, there is still a need for brighter electron sources to enable new and more demanding experiments to realize the full potential of the UED technique. Some examples of experiments that would require a higher average flux to complete in a reasonable time include mapping time-resolved phonon band structures from the weak thermal diffuse scattering\cite{Cotret2019} in low-density materials (e.g. 2D monolayers or bilayer heterostructures), performing time-resolved 3D reciprocal space mapping to resolve fine details of ordering dynamics in crystals, and observation of photochemistry in low-density liquid and gas-phase streams. In addition, improvement in generation of coherent electron beams with high average beam current will unlock new experimental possibilities which have thus far been limited to static electron microscopy, such as scanning electron nanodiffraction\cite{Zuo2004} and coherent diffractive imaging\cite{Elvio2018}.    

Here, we present the capabilities of a versatile setup for ultrafast electron scattering experiments at Lawrence Berkeley National Laboratory (LBNL) that applies a novel concept in electron generation via a continuous wave, RF photogun, producing high current electron bunches at relativistic energies up to MHz repetition rates. The instrument opens new avenues for UED with ultrabright electron beams with high flux and high repetition rates. In addition to the instrument characteristics, we provide a few case studies of the science that is being enabled by the Berkeley UED instrument.

\section{The High Repetition Rate Electron Scattering Instrument}
\label{section:beamline}
In this section, we describe the High Repetition-rate Electron Scattering (HiRES) instrument for the studies of atomically-resolved dynamics.

\subsection{Electron beamline}
The schematic layout of the HiRES beamline is provided in Fig \ref{HiRES-Beamline}. At its heart is the continuous wave, advanced photoinjector experiment (APEX) radio-frequency electron gun delivering high current (> nanoamperes) beams of relativistic electrons. The details of the state-of-the-art gun design which enable these unique characteristics have been discussed elsewhere \cite{PhysRevSTAB.15.103501}. 

Photoelectrons are generated from an easily exchangeable, high quantum efficiency, multi-alkali antimonide (\ce{CsK2Sb}) photocathode\cite{10.1063/1.3596450}, which has a much lower work function than other typical cathode materials. This allows for photoemission from visible photons that can be conveniently produced, e.g. by frequency-doubling the output of 1030 nm lasers, in contrast to generally more challenging and less efficient generation of UV photons required for metallic photocathodes. Emitted photoelectrons are extracted and rapidly accelerated to $\approx~750$~keV using high-gradient, RF fields ($\ge$ 20 MV m$^{-1}$; maximum driving RF power 120 kW CW). Following this, they pass through a 1.3 GHz CW, 2-cell RF bunching cavity (10 kW max power) which imparts a time-energy correlation for ballistic bunch compression at the sample. Optimal parameters of the bunching cavity depend strongly on the operating conditions such as initial pulse length and beam charge and are tuned to bring the electron bunch to a longitudinal focus at the sample plane a few meters downstream. After the RF compressor, bunches are directed through a pinhole aperture to filter out dark current generated by field emission due the high-amplitude RF. Various apertures featuring different sizes are available for this purpose and view screens are placed along the beamline to inspect the beam. Solenoid and quadrupole magnetic lenses are used to control the beam size during beam propagation. 

\begin{figure}[t!]
\centering
\includegraphics[width=8cm]{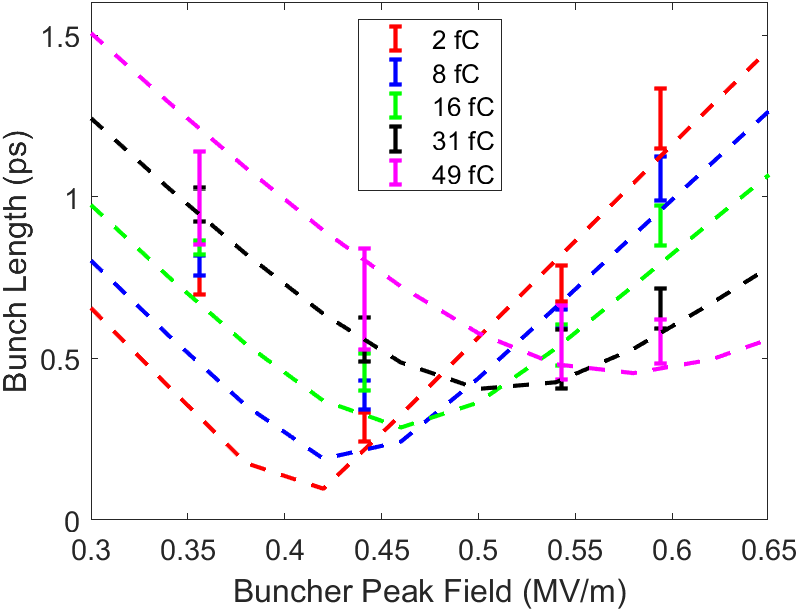}
\caption{Electron beam bunch length measured on a transverse deflecting cavity as a function of buncher cavity field.  Particle tracking simulation comparison is shown with dashed lines. }
\label{fig:bunching}
\end{figure}

After the buncher, HiRES branches out into diagnostic and UED beamlines as shown in Fig.~\ref{HiRES-Beamline}. The former corresponds to the straight line which has been developed for characterizing transversal and longitudinal properties of the electron bunches. It features a 1.3 GHz, 1-cell transverse RF deflecting cavity, which acts on the travelling electrons by providing a time-dependent momentum kick, mapping the longitudinal coordinates into the transverse plane (i.e. streaking). A calibrated detector placed downstream of the cavity records the streaked pattern from which direct measurement of pulse length and relative time of arrival can be made. Fig. \ref{fig:bunching} shows the measured electron bunch length as a function of the RF buncher field and beam charge extracted at the photocathode using the deflecting cavity. General Particle Tracer (GPT) \cite{GPT} simulations were carried out to compare with the measurements and generally show good agreement for beams longer than 500 fs, validating the beamline model. On the other hand, in the present configuration, although the model predicts down to 100 fs at 2 fC charges, direct measurements of sub-500fs pulse lengths were obscured mainly by the shot-to-shot time of arrival fluctuations \cite{cropp2023virtual}, which are sensitive to the buncher amplitude fluctuations near maximum compression. 

The second beamline is reserved for UED experiments. The path to the UED setup involves electrons being directed through the dogleg by a dipole magnet. In the process, the beam becomes dispersed due to the energy dependence of the bending radius in the dipole, which also allows for characterization of the bunch kinetic energy distribution. A triangular slit can be inserted here for energy collimation. A second dipole magnet then steers the electrons bunch towards the UED setup. Two quadrupole triplets are used to manipulate transverse beam properties and for optimization of probe beam shape and size at the sample and detector, respectively. The first one of these quadrupole triplets is located in the dogleg, and is generally used for dispersion compensation, such that the correlation between energy and transverse position after the dogleg, including at the sample, is suppressed. A second aperture is typically inserted after the second quadrupole triplet to further filter out dark current and spurious x-rays, and in some cases to collimate the beam to achieve a smaller spot size at the sample and/or detector.

The flexibility to tune different beamline parameters such as bunch charge, transverse/longitudinal beam properties and repetition rate allows for it to be adapted for a wide range of UED studies, providing access to materials and gas phase experiments as well as allowing to operate in nano-diffraction and projection imaging modes, in addition to the standard micro-diffraction mode. The range of electron beam parameters accessible at HiRES is summarized in Table~\ref{tab:electronParam}
\begin{table}[ht]
\caption{Typical operational parameter for HiRES.} 
\label{tab:electronParam}
\begin{center}       
\begin{tabular}{|l|l|} 
\hline
\rule[-1ex]{0pt}{3.5ex}  \textbf{Parameter} & \textbf{Value}  \\
\hline
\rule[-1ex]{0pt}{3.5ex}  Beam energy & 700-900 keV   \\
\hline
\rule[-1ex]{0pt}{3.5ex}  Bunch charge & $\ge10^{8}$ electrons per pulse \\
\hline
\rule[-1ex]{0pt}{3.5ex} Bunch length & 100-1000 fs (RMS)  \\
\hline 
\end{tabular}
\end{center}
\end{table}
\subsection{Laser System}
The HiRES beamline employs a commercial, Ytterbium-based femtosecond fiber laser system (Active Fiber Systems GmbH) for photoelectron generation and to serve as the excitation laser in pump-probe experiments. It outputs 315 fs FWHM pulses centered at 1030 nm (8 nm FWHM spectral bandwidth) with a tunable repetition rate (up to 250 kHz nominal, which also sets the repetition rate for UED experiments at HiRES), and a maximum pulse energy of about 200 \textmu J. The pulse energy rolls down above 50 kHz while the maximum average power of 50 W is maintained. The laser and the RF power amplifier for the electron gun are synchronized using a reference RF generator via a phase-locked-loop. HiRES is also equipped with an optical parametric chirped pulse amplifier (OPCPA) which accepts the output of the fiber laser and can provide extremely short laser pulses (approx. 10 fs FWHM) centered at 800 nm, opening the route for broadband pumping and generation of terahertz and mid-infrared light, further extending the tuning range for excitation. Typical parameters of the HiRES optical setup are reported in Table \ref{tab:laserParam}. 
\begin{table}[ht]
\caption{Typical laser parameters.} 
\label{tab:laserParam}
\begin{center}       
\begin{tabular}{|l|l|} 
\hline
 \multicolumn{2}{|c|}{Main Laser} \\
\hline
\rule[-1ex]{0pt}{3.5ex}  \textbf{Parameter} & \textbf{Value}  \\
\hline
\rule[-1ex]{0pt}{3.5ex}  Central wavelength & 1030 nm    \\
\hline
\rule[-1ex]{0pt}{3.5ex} Pulse energy & $\leq$100~\textmu J \\
\hline
\rule[-1ex]{0pt}{3.5ex} Pulse duration$^*$ & $\approx 315$~fs FWHM \\
\hline
\rule[-1ex]{0pt}{3.5ex}   Repetition rate & single shot $\rightarrow$  250 kHz  \\
\hline 
\rule[-1ex]{0pt}{3.5ex}  Beam spot size* & 250~-~1100  \textmu m  FWHM \\
\hline 
\rule[-1ex]{0pt}{3.5ex}  Accessible fluence range & 0.1~-~300 mJ cm$^{-2}$ \\
\hline 

 \multicolumn{2}{|c|}{OPCPA} \\
 \hline
 \rule[-1ex]{0pt}{3.5ex}  \textbf{Parameter} & \textbf{Value}  \\
\hline
\rule[-1ex]{0pt}{3.5ex} Central wavelength & 800 nm   \\
\hline 
\rule[-1ex]{0pt}{3.5ex} Pulse energy & $\approx$ 10\% of laser input \\
\hline 
\rule[-1ex]{0pt}{3.5ex} Pulse duration$^*$ & $\approx 10$~fs \\
\hline 
\end{tabular}
\end{center}
\end{table}
\subsection{UED setup}
As a pump-probe technique, UED requires a pump laser beam (typically in the optical regime) to prepare the system of interest out-of-equilibrium, and precisely synchronized electron pulses to take snapshots of the transient events in form of time delayed diffraction patterns. At HiRES, respective pump and probe beams are generated from a common laser (described above) which minimizes timing fluctuations to those of the electron transport. Further management of timing, including characterization of time-of-flight (TOF) jitters are carried out using virtual diagnostics that are described elsewhere \cite{cropp2023virtual, Filippetto_2016}. 
\begin{figure*}[t!]
\centering
\includegraphics[width=14cm]{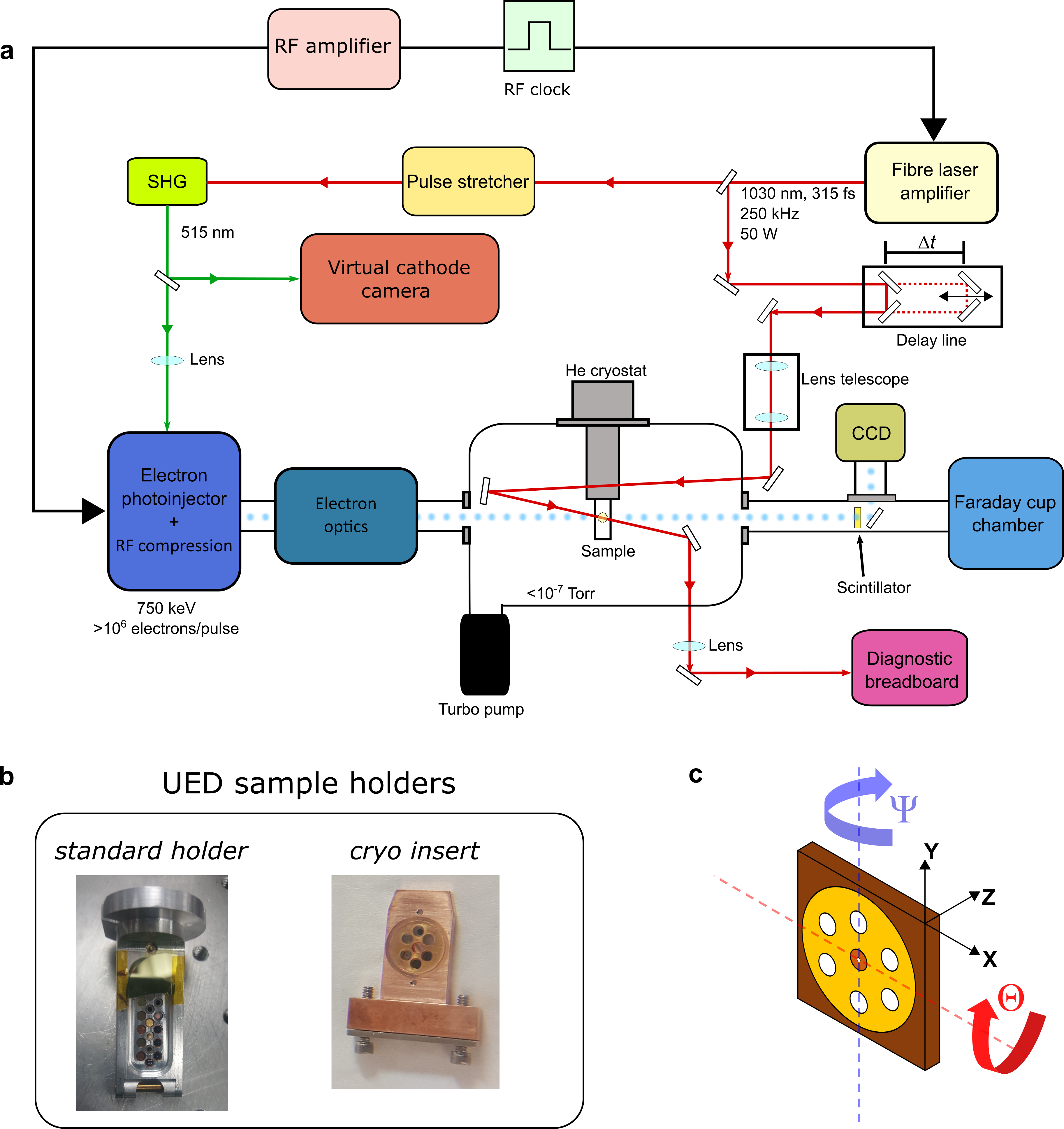}
\caption{(a) Schematic of the UED end station at HiRES (not to scale). The laser amplifier produces synchronized laser pulses used for generation of electron bunches and specimen pumping. (b) Choice of sample holders for materials science experiments depending on the temperature. (c) Illustration of accessible degrees of freedom using UED sample holders.}
\label{MS-UED}
\end{figure*}
As Fig \ref{MS-UED}a shows, the main laser output is split using a beam splitter to create the pump and probe arms. The higher energy pump beam is time-delayed with respect to the electron beam laser using a remotely controlled, mechanical delay line (which provides maximum possible delay of 1.5 ns), telescoped (dual-lens setup to adjust focal spot area at sample in range between 100 \textmu m$^2$ and 500 \textmu m$^2$ RMS) and sent to the sample chamber via one of the viewports. After entering the chamber, the pump beam reflects off a routing mirror and hits the target sample. The transmitted (residual) pump is harnessed by a mirror behind the sample holder, which directs the beam out through a side view port to the diagnostic breadboard containing a power meter and a beam profiler. The latter is setup to measure the beam spot size at the sample plane by virtue of 1:1 imaging as well as permitting real-time monitoring of beam position; the power meter provides an estimate of the beam pulse energy. This allows for a straightforward measurement of the fluence which can be adjusted by varying the pump power. 

The electron beam laser, on the other hand, is first sent into a stretcher, which elongates the pulse width of the laser from 315 fs to about 1 ps. The elongated pulses then are frequency doubled via a second harmonic generation (SHG) process in a beta-barium borate (BBO) crystal employing non-critical phase matching to output 515 nm pulses with photon energy matching the work function of the semiconducting, \ce{CsK2Sb} photocathode. The laser beam is then shaped by a mechanical aperture, and imaged onto the photocathode by a lens in near-normal incidence, front-illuminating geometry. The probe arm also features beam diagnostics containing a virtual cathode camera to determine beam size and profile at the cathode as well as fluence. 
\begin{figure*}[t!]
\centering
\includegraphics[width=17cm]{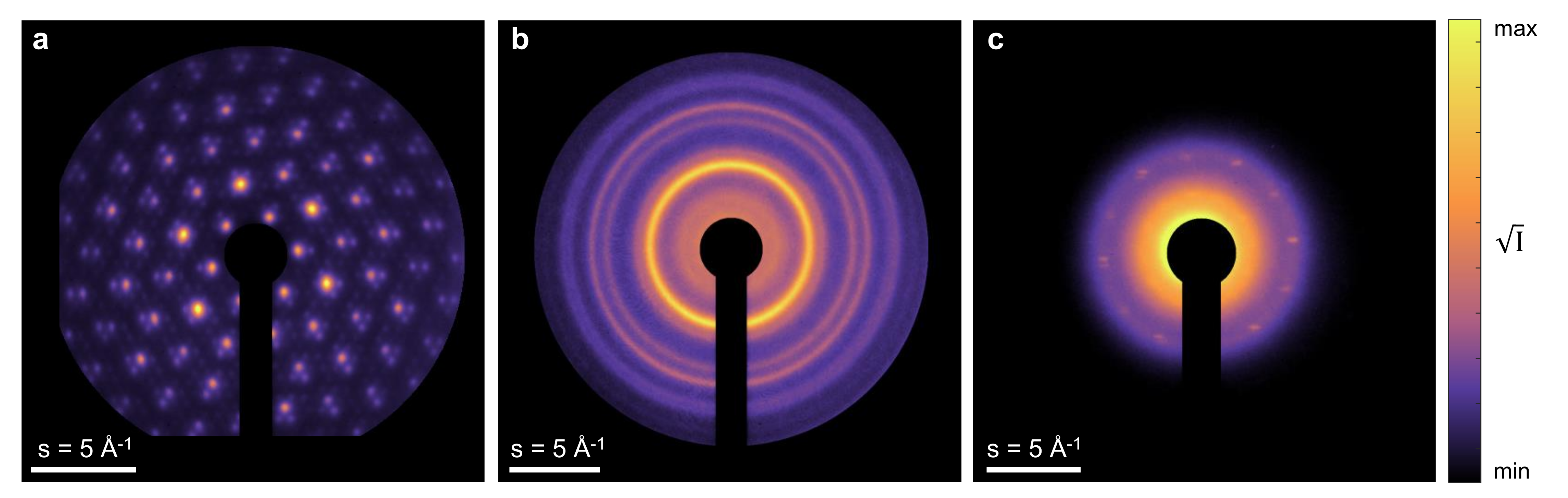}
\caption{Static electron diffraction patterns of (a) \ce{TaS2} [001], (b) polycrystalline gold, and (c) bicrystal graphene measured at HiRES using a repetition rate of 250 kHz. }
\label{fig:staticDPs}
\end{figure*}

Photogenerated electrons enter the sample chamber via the beam transport pipe, and travel straight downstream after interacting with the sample (crossing angle between the pump laser and electron probe is approximately $6 ^\circ$) and strike a cerium-doped yttrium aluminum garnet (Ce:YAG) scintillator screen, which emits visible radiation upon electron impact. The illuminated pattern on the back of the scintillator crystal is picked up by a mirror and imaged onto a 16-bit intensified charged-coupled-device (CCD, Princeton Instruments, PI-Max 4) orientated perpendicular to the beam path. Due to the near MeV-scale energy of the relativistic electron bunches, i.e. 750 keV, a detector distance of roughly 800 mm is used to capture the diffraction pattern. The unscattered electron beam is blocked by a beam stop to avoid detector saturation. Finally, the YAG screen and mirror can be retracted such that the beam is directed further downstream, where a Faraday cup is installed for measuring the beam current. The bunch charge needed in the experiment can be set by changing the laser power for electron generation or inserting pinholes in the beam path. 


The above is representative of the typical UED setup at HiRES. The end station can be purposed for the type of science that is of interest. Figure \ref{MS-UED}a depicts the sample chamber for materials science experiments.The salient feature of this format is a custom-engineered, in-vacuum cryogenic sample stage with four-axis motion, which is coupled via a thermal link to a commercial Gifford-McMahon cryocooler with low-vibration exchange gas interface. The sample stage includes a removable sample cassette (cryo insert) and is carefully shielded from impinging thermal radiation. Furthermore, it is strongly insulated from the motion stage's thermal mass using a Vespel bridge and layers of aluminized Mylar foil. As a result, sample temperatures in the range between 10 and 350 K are routinely achieved while retaining full four-axis positioning. Accessible temperatures below 20 K set the stage for structural dynamics research of certain classes of materials hitherto unstudied with UED, such as conventional superconductors.

Depending on the required temperature range, two different sample holders are available as shown in Figure \ref{MS-UED}b. The standard holder can accommodate up to 20 standard TEM samples including position reference apertures and YAG crystals to aid with alignment. This holder is primarily used for room temperature measurements. The cryo sample insert, on the other hand, has seven slots: the central one is typically reserved for an alignment pinhole for spatially overlapping pump and probe beams, one is typically left empty to allow the pump laser to pass through to exit the chamber for characterization, and the remaining five can be used for mounting samples. The front face of the holder has a holey (frosted) YAG plate with patterned laser-cut holes mounted on it providing access to the samples below. When moving between samples, the impinging beams can be viewed on the YAG surface to aid the alignment process. 

The holders are connected to the XYZ stage in the sample chamber which enables precise sample manipulation. As mentioned above, the cryogenic holder is strongly thermally decoupled from the motorized stages via the Vespel bridge to avoid raising it's base temperature. The standard holder, on the other hand, is magnetically attached directly to the XYZ stage which provides vertical and sideways translations (range >150 mm) as well as pitch ($\Theta$) and yaw ($\Psi$) rotations ($\pm 5^{\circ}$ on cryo inset and $> \pm 30^{\circ}$ when using standard holder) as shown in Fig \ref{MS-UED}c. Rocking curve and tilting experiments are, therefore, possible at wide range of temperatures on this setup.


\section{Examples of UED at HiRES} 
\label{section:SS}
\begin{figure}[t]
    \centering
    \includegraphics[width=7.5cm]{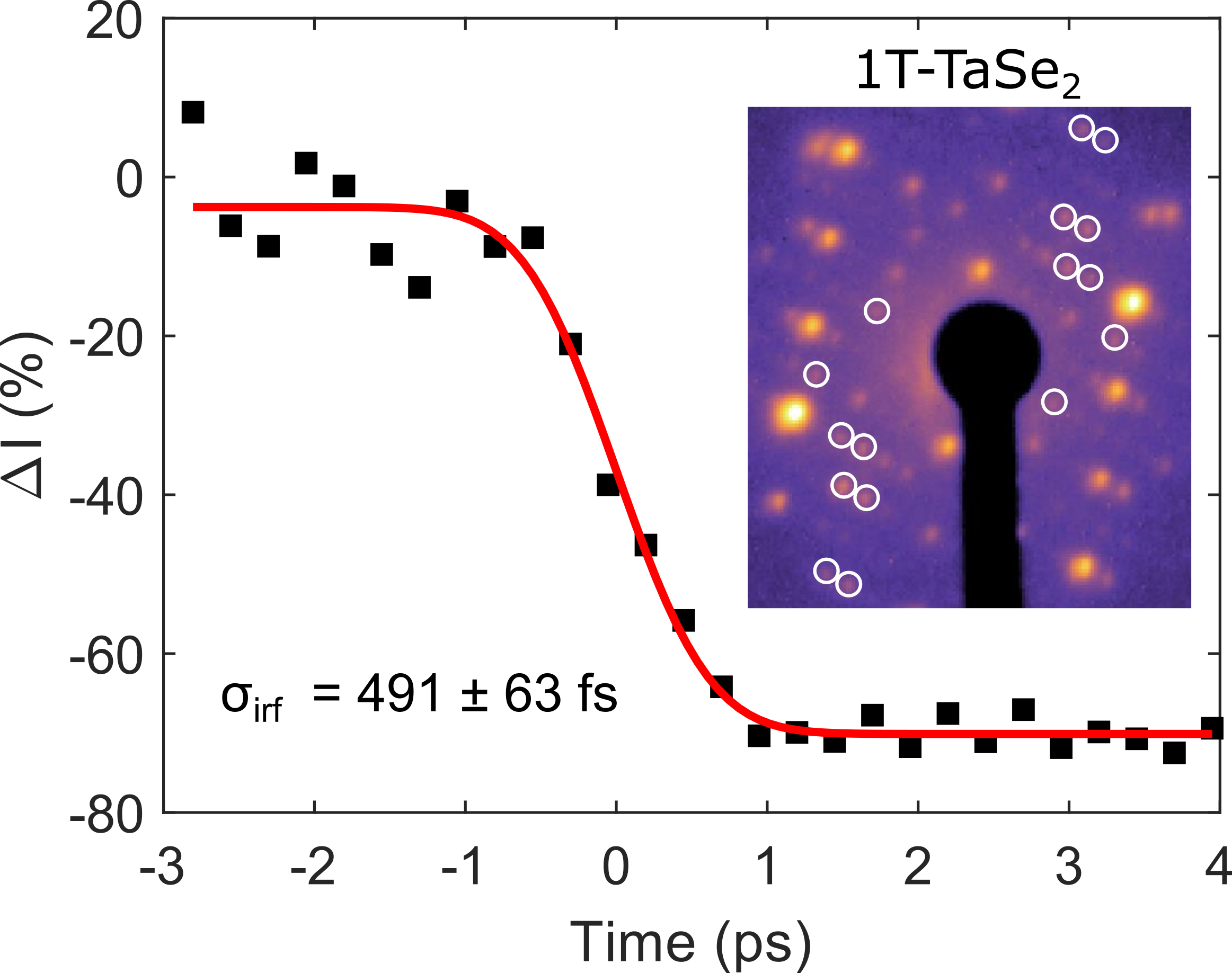}
    \caption{Characterization of time-resolution of pump-probe UED at HiRES using \ce{TaSe2}. The diffraction pattern is shown in the inset. The white circles denote the peaks that were chosen for analysis. An error-function fit (shown in red) has been used to determine the instrument response, characterized by RMS duration $\sigma_{irf}$. }
    \label{TaSe2}
\end{figure}

Materials science experiments are a major focus at HiRES UED \cite{Siddiqui2021,Durham2022} and in this section, we provide examples of the types of measurements that are possible. Fig.~\ref{fig:staticDPs} features static electron diffraction patterns obtained from selected solid-state samples, which represent types of materials which are of frequent interest for UED studies including quantum materials\cite{Wang2022,JLi2022}, polycrystalline metals \cite{Sciaini2009,Qi2022,Tauchert2022}, and 2D materials \cite{Sood2023,Gulde_2014}. As can be seen, a large momentum transfer, \textit{s} range (where $s = \frac{2\pi }{\lambda }\sin \left ( \frac{\theta }{2} \right )$, $\lambda$ is the electron wavelength and $\theta$ is the scattering angle) of $\pm10$~\AA$^{-1}$ is accessible which allows to track the dynamics of several diffraction orders simultaneously. In addition, momentum resolution as small as 0.1 {\AA}$^{-1}$ RMS has been achieved so far. The high flux of electrons also permits acquisition of high fidelity diffraction patterns from weakly scattering materials, such as monolayer graphene as shown in panel c of Fig.~\ref{fig:staticDPs}. Under the present conditions, the temporal resolution of the setup was demonstrated using a thin flake of \ce{TaSe2} \textemdash~a layered charge density wave (CDW) material, by observing the photoinduced suppression of the CDW peak caused by 1030 nm laser excitation at 1.9 mJ cm$^{-2}$ as shown in Fig~\ref{TaSe2}. The electron bunch charge at the sample was approximately 2 fC ($1.2 \times 10^{4}$ electrons/pulse) for this measurement and an error fit to the data recovered an instrument response of sub-500 fs RMS. This makes it possible to study a range of phenomena including electron-phonon couplings, charge density wave dynamics, and coherent lattice motions.   

\subsection {Ultrafast heating in single crystal gold} 
\begin{figure*}[t!]
    \centering
    \includegraphics[width=17cm]{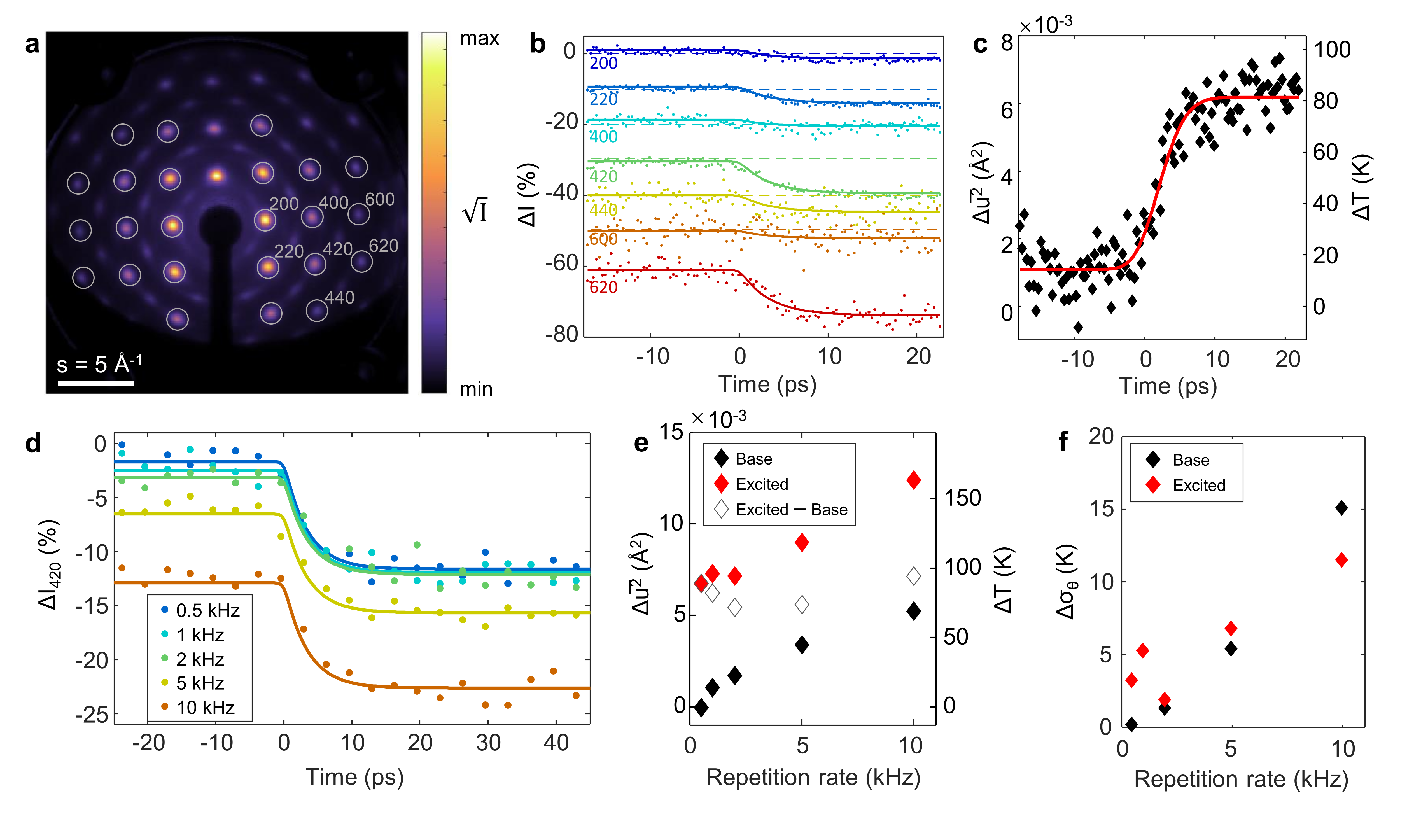}
    \caption{Repetition rate effects on lattice temperature dynamics during pump-probe UED of freestanding gold foils (a). Diffraction from the [001]-oriented single crystal gold film (b). Temporal evolution of Bragg orders. (c). Time-resolved lattice temperature extracted using dynamical scattering simulations (black diamonds) fit with a two-temperature model convolved with the instrument response (red line). (d). Temporal evolution of the 420 order at varying repetition rates. (e). Temperature rise directly following pulses (Excited) and residual temperature (Base) for varying repetition rate. (f). Change in sample rippling extracted for the same conditions.}
    \label{Au_TimeResolved}
\end{figure*}
We demonstrate the capability to perform relativistic UED at kHz repetition rates on freestanding thin films using single-crystal gold foil as example. Photoinduced carrier dynamics in gold have been well characterized by many techniques, including UED, making this an ideal benchmark system. For the experiment, we excited commercially-available 11 nm [001]-oriented Au films with 1030 nm laser pulses at 3.9 mJ cm$^{-2}$ peak fluence and recorded the time-resolved electron diffraction patterns to measure the lattice temperature dynamics. The pump laser spot size was about 700 \textmu m FWHM in effort to provide mostly uniform excitation over the 350 \textmu m FWHM electron probe size.

A static electron diffraction image from the Au film is shown in Fig~\ref{Au_TimeResolved}a, revealing a four-fold symmetry suggesting that the film was oriented near to the [001] zone axis. The repetition rate for this measurement was set at 0.5 kHz and each electron pulse contained $10^4$ particles on average. Several Bragg orders up to and including 620 are clearly resolved, allowing simultaneous analysis of multiple diffraction peaks, improving the accuracy of lattice temperature retrieval. 

Fig~\ref{Au_TimeResolved}b shows a plot corresponding to the time-dependent intensity change of selected families of Bragg peaks together with global exponential fits to the time-traces to clarify the observed response. The suppression of the Bragg peaks occurs due to transfer of thermal energy from the electron subsystem to atomic vibrations via electron-phonon coupling causing an increase in the lattice temperature. 

In conditions where the kinematical scattering condition holds, the thermal suppression of diffraction peaks can be well described using the Debye-Waller Model (DWM)  via the following relation
\begin{equation}
    -\log{\frac{I}{I_{0}}} = s^{2} \frac{\overline{u^2}}{3}
\end{equation}
where $\overline{u^2}$ corresponds to  mean square atomic displacements. The DWM predicts that the magnitude of the intensity change increases with the scattering vector, $s$. However, our data show deviation from this behavior attributed to non-negligible multiple scattering, which calls for models that account for dynamical effects. Such a model was developed in-house\cite{Durham2022} and applied to the data to extract the time-dependent lattice temperature following photoexcitation. The obtained results are presented in Fig~\ref{Au_TimeResolved}c alongside a two-temperature model (TTM) fit. An electron-lattice coupling constant, $G = 2.6 \times 10^{16}$~Wm$^{-3}$K$^{-1}$ was employed and the fit was convolved with the instrument response which shows good agreement with the data validating the utility of the model. As can be seen from the figure, the residual temperature rise is less than 20 K for measurements done at 0.5 kHz repetition rate.    

Following this, we tested the response of the sample at increasing repetition rates to investigate the onset of residual heating and the damage threshold. For this, we carried out experiments on the same sample with repetition rates of 1, 2, 5 and 10 kHz. The time-resolved intensity changes of the 420 order from these measurements are illustrated in Fig \ref{Au_TimeResolved}d. The dynamics do not appear to be significantly affected by the increased repetition rate up to 10 kHz. However, a negative offset in the peak intensity characteristic of accumulated heating is observed and increases with repetition rate. 

We can quantify the lattice temperature before and after the arrival of the laser pulse at each repetition rate by once again applying the dynamical scattering model to the first seven diffracted orders. These temperatures were extracted by averaging the data recorded at time delays ranging from -24 ps to -6 ps (i.e. pre time zero, denoted ``base'' temperature) and from 13 ps to 47 ps (i.e. post time zero, denoted ``excited'' temperature), and then applying the model to extract the lattice temperature rise and change in RMS tilt spread. The results are shown in Figure~\ref{Au_TimeResolved}e-f. Indeed, at higher repetition rates, as expected accumulated heating leads to an increase in both base and excited temperatures, but the rise due to each individual pulse (``excited $-$ base'') remains roughly the same. For this experimental configuration, the change in base temperature when using 5 kHz repetition rate is found to be more than half the temperature rise due to the individual pulse, and at 10 kHz the base temperature rise is nearly equal to that imparted by an individual pulse. An attempt to further increase repetition rate to 25 kHz damaged the sample: irreversible melting and recrystallization occurred within seconds. 

These results show that heat accumulation must be carefully considered to access the benefits of higher repetition rates for UED of large films, even for materials with high thermal conductivity such as metallic foils. In this experiment, the 11 nm foil was supported by a gold TEM grid with 20 \textmu m wide grid bars and 60 \textmu m wide holes excited using a 700 \textmu m wide laser spot: the very high aspect ratio, broad excitation region, and operation in vacuum makes heat dissipation slow. Under high repetition rate operation, the base temperature will be especially high at the centers of the windows as well as the center of the overall laser spot: this inhomogeneous temperature distribution complicates data interpretation and also limits the excitation fluence that can be used before damaging the sample. Improvements to sample platform design, including the geometry and materials used, should be examined. However, UED experiments impose many requirements which need to be considered including sufficient electron transparent area over the probed region, access to varying tilt angles, and operation in vacuum. 

On the other hand, the benefits of high repetition rate are much clearer for smaller samples. In this case, the pump and probe lateral size can be reduced, and heat dissipation can be much faster since the maximum usable repetition rate scales quadratically with the pump diameter\cite{filippetto2022rmp}. Indeed, repetition rates approaching MHz have been utilized in ultrafast transmission electron microscopes: in one example, the pump and probed area were reduced to less than 2 \textmu m by depositing a thick gold mask with holes over the sample mounted on a silicon nitride support\cite{danz2021ultrafast}. Many types of samples of interest for UED study are limited to several micron sizes due to challenges of sample preparation, including some exfoliated 2D material flakes, quasi-1D material wires, and thin lamellae prepared using focused ion beam milling. In these cases, higher repetition rates will be accessible and will provide a significant improvement in signal. In the nanoscale limit, high repetition rate becomes essential to record diffraction patterns with sufficient signal. In the next section, we demonstrate the utility of HiRES in the nanoscale regime.

\subsection{Nanoscale UED}

Probing ultrafast dynamics in nanoscale volumes is an important emerging research area. Heterogeneous nanoscale dynamics may influence the overall behavior of ultrafast processes, as suggested by results from prior UED studies which hypothesized grain-dependent structural phase transformations\cite{otto2019vo2} and transient local defects\cite{gedik2019defects} were occurring within the probed area. Low-dimensional nanocrystals likely host unique, size-dependent photoinduced dynamics and transient structures due to quantum confinement, as they are known to have altered electronic and lattice modes. The ability to perform UED at the nanoscale could reveal heterogeneous dynamics and access novel phenomena at quantum length scales \cite{danz2021ultrafast,Johnson2023}.

However, reducing the probed area from micrometers to nanometers demands much higher brightness of the electron probe. For a given \textit{s} resolution set by the RMS angular spread, $\mathrm{\sigma_{\theta}}$, the maximum achievable diffracted intensity I(\textbf{s}) is related to the material's intrinsic electron scattering probability S(\textbf{s}), the probed sample area, $\mathrm{A_{sample}}$, and the average transverse beam brightness at the sample, $\mathrm{B_{4D,av}}$:
\begin{equation}
    I(\mathbf{s}) \propto S(\mathbf{s})B_{4D,av} \sigma_{\theta}^{2} A_{sample}
\end{equation}
Due to low transverse brightness, typical sub-kHz MeV-scale UED beamlines operate with tens of \textmu m or larger beams and examine thin film samples of similar size. Now that the HiRES beamline can access kHz to MHz repetition rates, the average brightness is increased by orders of magnitude, permitting MeV-scale UED probing of nanoscale regions.
\begin{figure}[t]
    \centering
    \includegraphics[width=8cm]{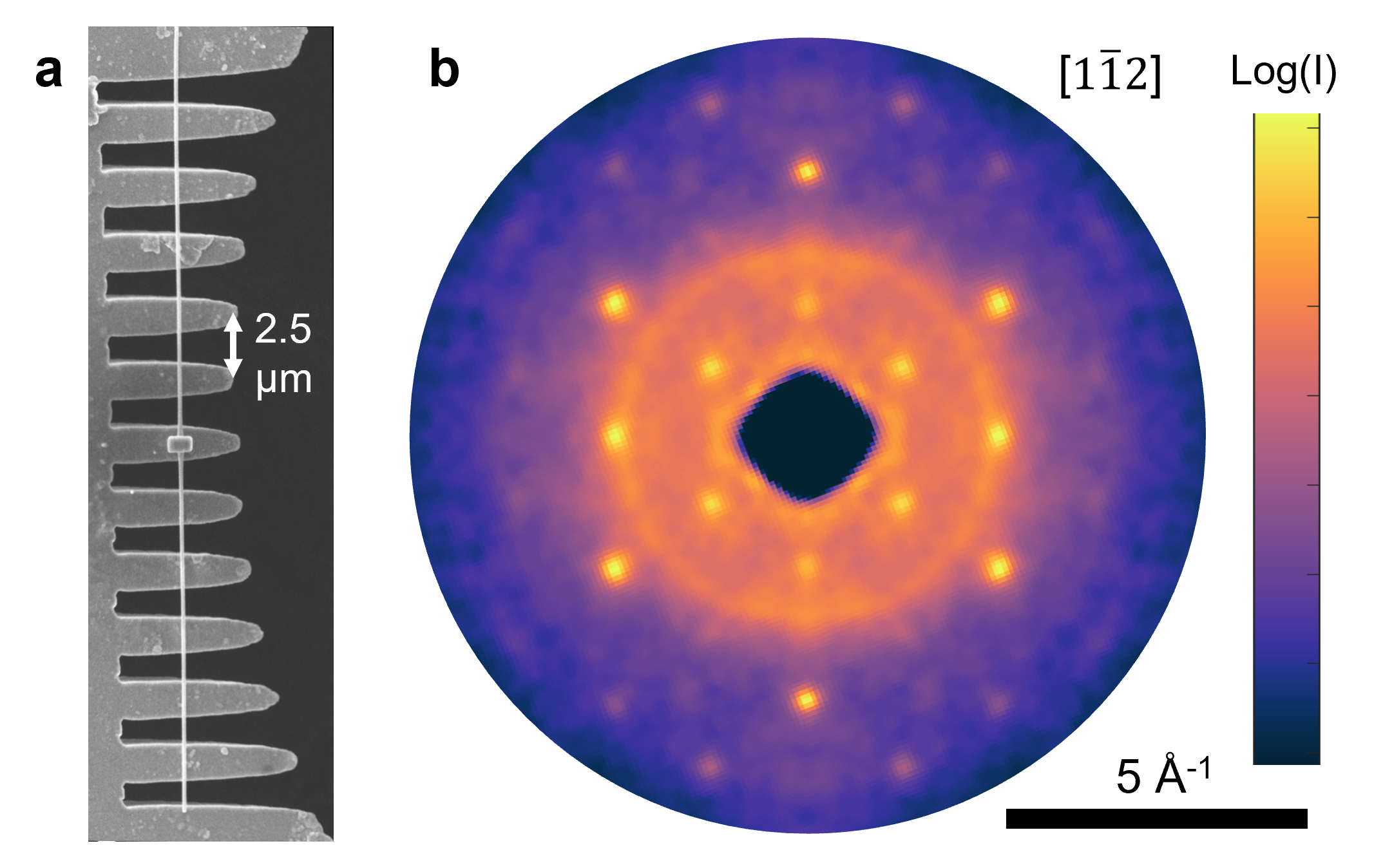}
    \caption{UED pattern from an individual \ce{VO2} nanowire at HiRES (a) SEM image of the prepared sample on a fabricated Cu microcomb support (b) Symmetrized diffraction pattern from the monoclinic $[1 \overline{1} 2]$ zone axis.}
    \label{fig:VO2}
\end{figure}

\begin{figure*}[t!]
    \centering
    \includegraphics[width=16cm]{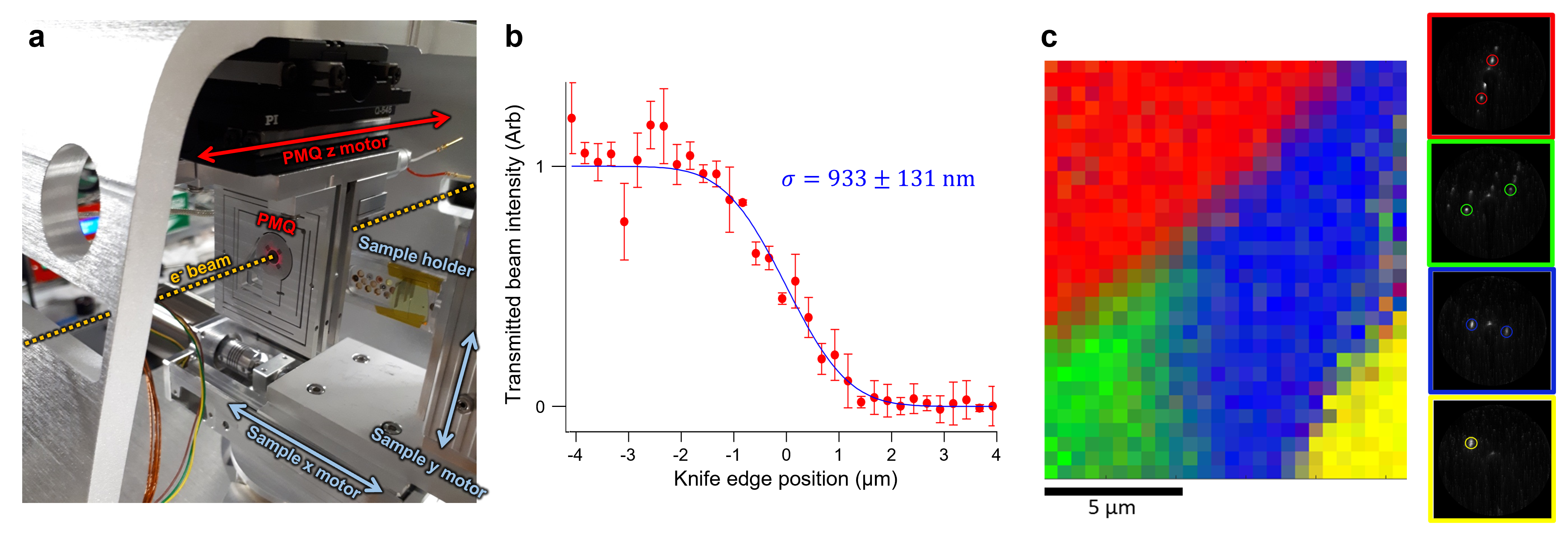}
    \caption{Ultrafast electron nanodiffraction (nano-UED) at HiRES (a) Experimental chamber setup including a PMQ triplet on a motorized stage to focus electron bunches at the sample. (b) Knife-edge scan of a focused beam at the sample plane showing sub-\textmu m spot size. (c) Two-dimensional nanodiffraction mapping of grain boundaries in a polycrystalline Ti-Al alloy. The RGB color values are weighted by the fraction of the signal corresponding to the circled peaks in the four individual grain patterns shown at right.}
    \label{fig:nanoUED}
\end{figure*}
One approach to nanoscale probing is to illuminate a nanoscale sample with the full microscale UED beam. As an initial demonstration, we measured a diffraction pattern from a single \ce{VO2} nanowire on a custom Cu microcomb support fabricated using focused ion beam milling. An SEM image of the wire is shown in Figure \ref{fig:VO2}a. It has a square cross section with about 150 nm side length. The wire was placed on the support with a micro-manipulator and secured using e-beam deposition of a platinum patch at the center. A diffraction pattern recorded along the $[1 \overline{1} 2]$ zone axis of the monoclinic phase is shown in Figure \ref{fig:VO2}b. Here, we averaged 10 frames integrated for 8 seconds each while using 50 kHz repetition rate. Several \ce{VO2} diffraction peaks are resolved, distinguishable from the weak diffraction ring background from the polycrystalline Cu.
A second approach is to focus the electron probe to nanoscale dimensions at the sample and perform ultrafast electron nanodiffraction (nano-UED). We demonstrated this capability using a custom permanent magnet quadrupole (PMQ) triplet lens as shown in Figure~\ref{fig:nanoUED}a. By scanning the beam with a knife edge milled into a 75 nm gold film deposited on a 30 nm $\mathrm{SiN_{x}}$ membrane, we measured sub-\textmu m spot sizes at the sample like shown in Figure~\ref{fig:nanoUED}b. Using this focused beam, we performed a scanning nanodiffraction measurement and mapped the local grain structure in a polycrystalline Ti-Al wedge, showing the ability to locate individual grains and interfaces in a material with nanoscale precision (Figure~\ref{fig:nanoUED}b). Each pattern was recorded in about 5 seconds at 250 kHz repetition rate. More detailed characterization of the nanofocused beams and further details and demonstrations of the nano-UED capability are reported elsewhere\cite{ji2019ultrafast,ji2019knife}.

These two approaches provide different advantages and are better suited for different types of samples: the full beam is well suited for studying average dynamics in high aspect ratio structures like nanowires or large ensembles of nanostructures like nanoparticles, where a focused beam would only sample from a small region and provide limited signal. On the other hand, a nano-focused beam provides distinct advantages and access to information not obtainable with the full beam. For instance, the focused beam can be used to sample and compare various nanoscale regions within the same sample to identify and map heterogeneity. It can also reduce background contributions from membranes or other support structures, like those observed in our example nanowire pattern (Figure~\ref{fig:VO2}b). Either approach requires collimation of the beam to reduce the probe to nanoscale size: when sending the full beam, useful scattering is only obtained from the portion of the beam illuminating the sample such that $\mathrm{A_{sample}}$ is set by the sample size, while when producing a nanofocused beam, the beam is significantly apertured upstream to reduce the 4D emittance such that $\mathrm{A_{sample}}$ is set by the small focal spot of the beam. 

There are still challenges to overcome in the sample preparation for such nanocrystals, as they can be more fragile and mobile than films. Sample mounting like shown in Figure~\ref{fig:VO2}a is adequate for static measurement, but for pump-probe experiments involving large dynamic strains, other designs may be needed. For instance, one end may need to be left free or a freestanding nanobeam may be secured at both ends and pulled taught using a MEMS device to accommodate dynamic strains along the wire length. Thermal management also needs to be considered, such as using masks to reflect spurious laser power and developing sample platforms with improved heat sinking.

Altogether, these initial measurements demonstrate the potential for performing MeV-scale UED experiments at the nanoscale at HiRES. Further optimization of the beam parameters and development of sample platforms will allow study of smaller and more diverse kinds of materials and microstructures, ultimately providing a route to examine femtosecond structural dynamics in regimes of quantum confinement.

\subsection{Gas-phase UED at HiRES} 
Studies of isolated quantum systems are of crucial importance for addressing the nature of chemical bonds and the transition state \cite{ZewailBook94}. Chemical transformations (e.g. bond dissociation, isomerization etc) occur on ultrafast time scales between ten femtoseconds to a few picoseconds, and often involve complex pathways \cite{Warren1993,Deb2011}. Disentangling the structural evolution of isolated molecular species in real time is the primary goal of gas phase ultrafast electron diffraction (GUED) \cite{Shen2019}. Insights from GUED experiments can reveal key information on how energy is partitioned into reactive modes that direct chemistry \cite{MillerAndIschenko2017,Ihee2001}, the role of conical intersections \cite{Yang2018}, and how introducing a (solvent) bath could affect chemical dynamics \cite{Pigliucci2007,Underwood2003,Kwak2006}. 

Gas-phase UED studies raise numerous technical challenges related to signal-to-noise and, in case of time-resolved measurements, temporal broadening due to group velocity mismatch between the dynamics-initiating laser pulse and electron probe \cite{Williamson1993}. The latter is circumvented by using relativistic electrons that have comparable velocities to that of the speed of light. However, the orders of magnitude lower density of scatterers when compared with solid state specimens places stringent requirements on flux needed to reach signal levels well above the noise floor to permit pattern inversion. While great strides have been made in the field to overcome these issues from the introduction of novel analysis methods, such as ratio method \cite{Burgi_1989,Ischenko1994,Maggard1995}, laser alignment of molecules \cite{Hensley2012}, and proposed use of direct electron detectors \cite{Vecchione2017}, most of the work has been limited to low electron flux to avoid space charge broadening and low-repetition rates leading to very lengthy experiments that lead to hardware drift that hamper the signal-to-noise. The promise of HiRES to this end is to provide an order of magnitude improvement in signal-to-noise thanks to very high-repetition rates and high-current density electron bunches. This section reports on proof-of-concept trials for gas-phase UED experiments at HiRES

\subsection*{HiRES-GUED setup}
\begin{figure}[t]
\centering
\includegraphics[width=0.45\textwidth]{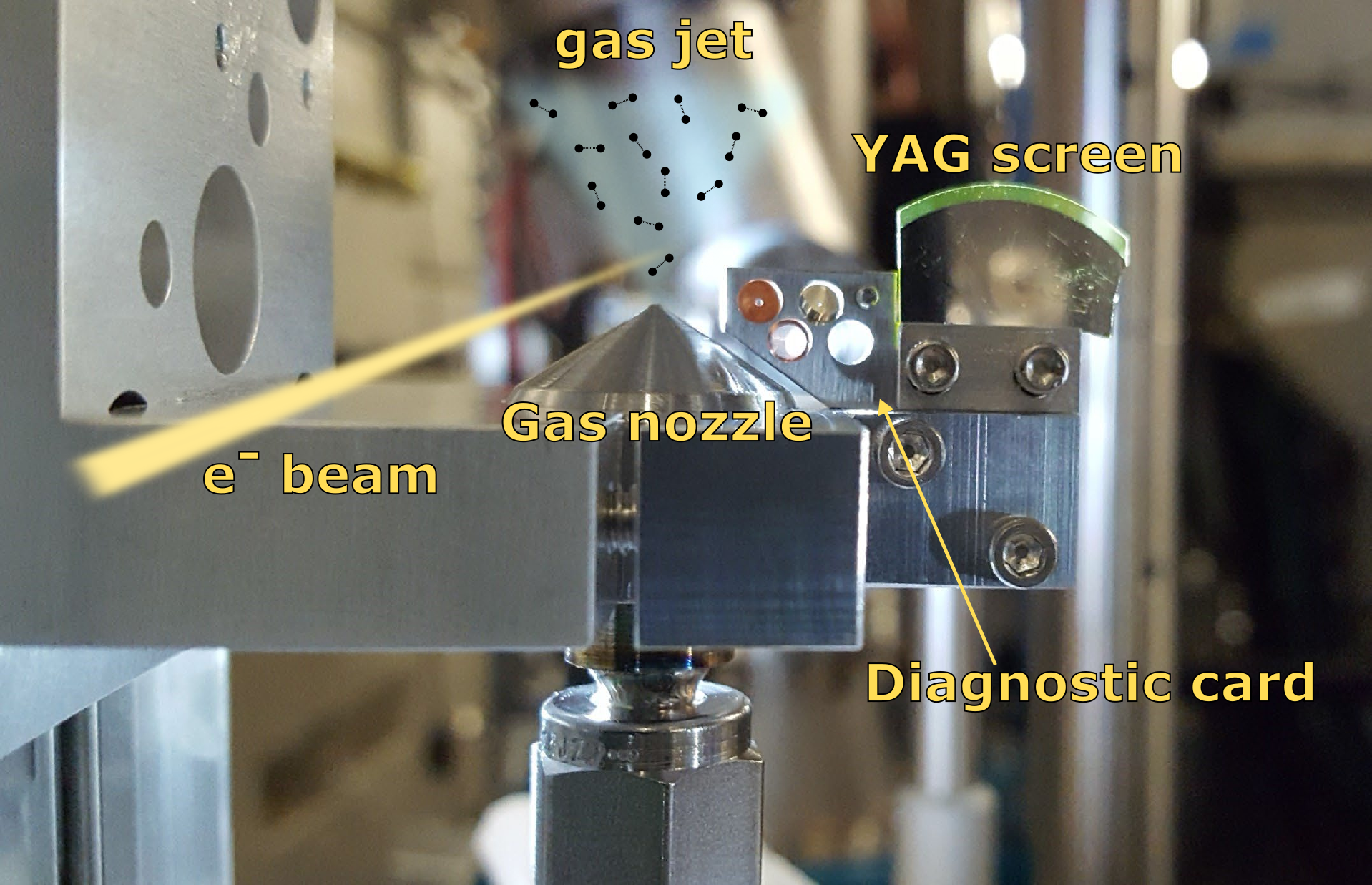}
\caption{Overview of the gas-phase electron diffraction sample delivery system at HiRES.}
\label{GUEDSetup}
\end{figure}

The GUED setup comprises of a compact, custom-built gas delivery and alignment assembly which is mounted onto the translation stage as shown in Fig~\ref{GUEDSetup}. It consists of a gas nozzle connected to a gas line to allow delivery of target molecules into the sample chamber. The inlet diameter of the nozzle measures at 21~\textmu m whereas the outlet has a diameter of 32~\textmu m. Teflon or metal tubings are used for gas transport and the backing pressure can be adjusted from the outside using pressure regulators. The target gas is delivered in form of a continuous, supersonic  beam and the pressure inside the chamber is maintained by turbomolecular pumps of large pumping capability (> 800 Ls$^{-1}$). For alignment purposes, a diagnostic card is attached to the nozzle assembly and holds two pinholes of diameter 100~\textmu m and 50~\textmu m respectively, a TEM grid and a small piece of YAG screen. The YAG screen and pinholes are used as position references for overlapping the gas jet and electron beam by translating back and forth between them. Diffracted electrons impinge on a YAG scintillator located downstream of the GUED setup assembly at a distance of about 597 mm from the gas jet and the illuminated pattern on scintillator is captured by the CCD camera. 

\subsection*{GUED of nitrogen gas }

\begin{figure*}[ht!]
\centering
\includegraphics[width=17cm]{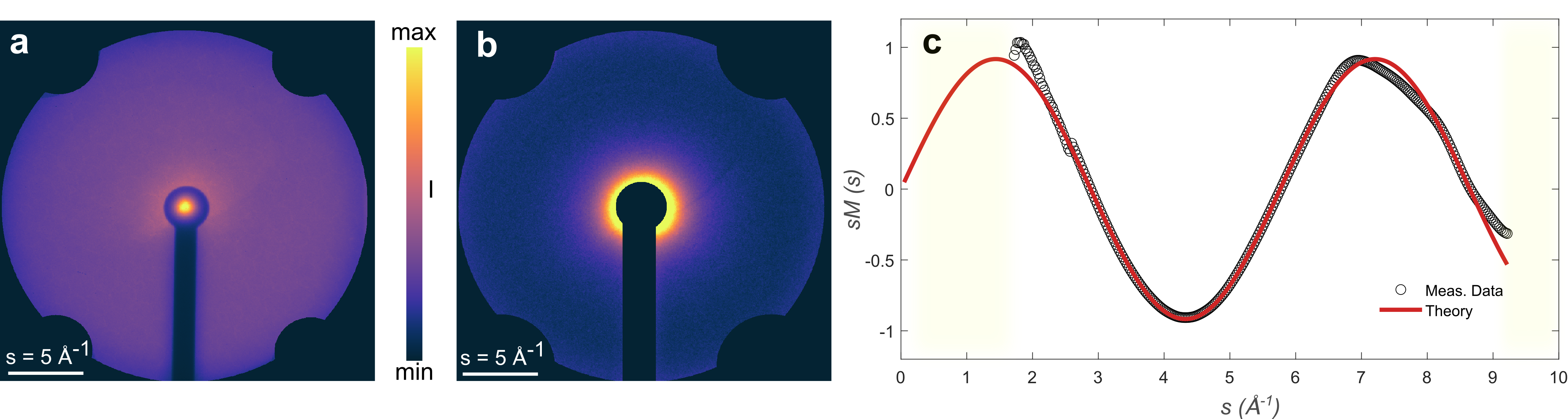}
\caption{(a) Raw electron diffraction pattern of nitrogen gas acquired with one second acquisition (b) Processed diffraction signal after intensity normalization and background subtraction. The solid line is the theoretical prediction using the Independent Atom Approximation (c) Modified scattering, $sM$~obtained from diffraction signal in (b). Regions below 1.5~\AA$^{-1}$ and above 9.5~\AA$^{-1}$ have been masked (shaded regions). }
\label{Fig7:GUEDPatterns}
\end{figure*}

We demonstrate the capability of the HIRES GUED setup to capture a diffraction pattern employing very short acquisition times using nitrogen (\ce{N2}) as the target system. Nitrogen gas was introduced with a backing pressure of 360 Torr resulting in pressure of $1.2\times10^{-4}$ Torr inside the sample chamber. Electron bunches at 250 kHz repetition rate with average beam current of 2.3 nanoamperes (nA) corresponding to $5.8\times10^4$ electrons/pulse were intercepted by \ce{N2} in the gas jet and undergo scattering onto the YAG screen. 

The raw pattern of \ce{N2} diffraction is shown in Figure \ref{Fig7:GUEDPatterns}a which was recorded for a total acquisition time of one second. As indicated on the figure, approximately 10~{\AA}$^{-1}$ is accessible which is sufficient for pair-distribution function analysis of most chemical systems in gas phase \cite{Centurion2022}. As expected, the scattering intensity is highest at low momentum transfers, \textit{s} near the center but drops rapidly towards large momentum transfers. The edge of the detector screen is obscured by four camera port holding screws and is masked. The figure also shows the leak of the non-diffracted beam through the beam block which can conveniently be used for intensity normalization. The diffraction pattern after background image (taken with the gas jet turned off) subtraction and normalization is presented in Figure \ref{Fig7:GUEDPatterns}b and shows a high-quality gas phase diffraction pattern  measured at a fraction of acquisition times of several minutes used in many GUED experiments and from which structural information can be readily obtained. To put this claim on a more solid footing, we extract the scattering intensity from the diffraction pattern. This is achieved by first radially averaging the isotropic patterns from the center and noting that the total scattering intensity ($I_T$) is a sum of contributions from atomic and molecular scattering (denoted $I_A$ and $I_M$, respectively) as a function of momentum transfer, i.e.

\begin{equation}
I_{T}(s) = I_{A}(s) + I_{M}(s)
\end{equation}

The $I_A$ term depends only on the scattering amplitudes of individual atoms whereas the $I_M$ term contains structural information coming from the interference between the scattered electrons by different atoms in the molecule. The latter is better visualized as modified scattering, \textit{sM (s)} defined as 

\begin{equation}
sM(s) = s\frac{I_{A} (s)}{I_{M}(s)}
\end{equation}

Following well-established data analysis routines \cite{Centurion2022}, the experimental \textit{sM} for \ce{N2} was extracted from diffraction pattern in Figure \ref{Fig7:GUEDPatterns}b and compared with theoretically calculated \textit{sM} \cite{Yang2016} as shown in Fig \ref{Fig7:GUEDPatterns}c. The excellent agreement between the experimental and theoretically calculated \textit{sM} of nitrogen gas validate the use of HiRES as an emerging UED setup for gas phase studies rivaling the best UED instruments in operation today by providing many fold improvements in reducing acquisition times through its very high repetition rates, high-brightness and proportionally high gain in signal-to-noise. For example, the above example was enabled by nearly $10^3$ times higher average current than state-of-the-art relativistic gas UED instrument\cite{Shen2019}. Future work aims to explore time-resolved experiments to benchmark the performance of HiRES-GUED with well-studied systems such as cyclohexadiene (CHD).      
\section{Outlook and Conclusions} 
\label{section:conclusions}
Rapidly evolving electron source technology has already enabled new routes for investigation of the structure-function relationship in a wide class of materials relevant to chemistry, biology and physics. However, to take the next leap forward and explore new vistas in ultrafast science using UED, electron sources with high beam brightness, high-flux and high repetition rates are greatly desired. In this article, we have highlighted such a source at LBNL in HiRES which combines very high repetition rates up to megahertz, high beam flux with the possibility of $\ge 10^{8}$ electrons per pulse, and relativistic energies of 750 keV. 

So far, thanks to the large instrument parameter space, we have demonstrated several experimental modes at HiRES beyond typical micro-diffraction of materials, including gas-phase diffraction, scanning nanodiffraction, and even projection imaging\cite{ji2019ultrafast}. Even still, we anticipate the high average brightness of relativistic electrons can be exploited for additional new modes. For instance, real-space imaging might be achieved via coherent lens-less imaging modes or by installing an objective lens. 
%
%
The combination of relativistic electron bunches with a cryogenic stage may also enable access to biological specimens such as proteins and viruses, which may benefit from low dose-rate or, on the other hand, single-shot ``diffract and destroy'' approaches for imaging and diffraction \cite{Fitzpatrick_2015,Fitzpatrick_2013,Flannigan_2010,Henderson_1995}.

The access to high repetition rates will also be an indispensable asset for future gas and liquid phase studies as it will widen the palette of systems that could be studied with UED to include those that have low vapor pressures or low scattering power. To this end, the parallel advances in laser technology providing high powers at high repetition rates complement the use of high repetition rate electron sources well. However, to take advantage of this feature in the case of solid materials will require strict thermal management as discussed earlier. Future work could investigate strategies including sample mounting, reduction of laser spot size, and placing a metallic mask on the sample outside of the region of interest.  

Finally, the instrument temporal resolution is presently limited by the pulse length of the excitation laser used (approx. 315 fs), and by the   amplitude and phase fluctuations of the accelerating and bunching RF fields. The large distance between the photocathode and the experimental chamber (due to space constraints) increases the sensitivity of the final time jitter to beam energy, leading to very tight stability requirements. Work towards 100 fs resolution is ongoing and includes the use of the OPCPA beam for experiments, the shortening of the UED line, the installation of an alternative chamber on the straight line, and the development of non-invasive time-stamping techniques\cite{cropp2023virtual}. These upgrades are expected to enhance the versatility of the HiRES-UED instrument and enable a host of novel experiments that could unveil new domains of structural dynamics research and beyond.
\begin{acknowledgments}
This material is based upon work supported by the U.S. Department of Energy (DOE), Office of Science, Office of High Energy Physics and Office of Basic Energy Sciences, Division of Chemical Sciences Geosciences and Biosciences  under contract number 89233218CNA000001 and DE-AC02-05CH11231. K.M.S., D.F., and R.A.K. acknowledge support through the Laboratory Directed Research and Development (LDRD) Program of Lawrence Berkeley National Lab under U.S. Department of Energy (DOE) Contract DE-AC02-05CH11231. F.C. acknowledges support from NSF PHY-1549132, Center for Bright Beams. F.C. also acknowledges support from the U.S. Department of Energy, Office of Science, Office of Workforce Development for Teachers and Scientists, Office of Science Graduate Student Research (SCGSR) program. The SCGSR program is administered by the Oak Ridge Institute for Science and Education for the DOE under contract number DE‐SC0014664. D.F. and S.P. acknowledge support for Machine learning studies at HiRES by the Laboratory Directed Research and Development program of LBNL under U.S. DOE Contract DE-AC02-05CH11231. D.B.D. and A.M.M. acknowledge support from STROBE: A National Science Foundation Science and Technology Center under Grant No. DMR 1548924.  J. L and J. W. acknowledge support by U.S. NSF Grant No. ECCS-1953803. Work at the Molecular Foundry was supported by the Office of Science, Office of Basic Energy Sciences, of the U.S. Department of Energy under Contract No. DE-AC02-05CH11231.
\end{acknowledgments}
\section*{Data Availability Statement}
The data that support the findings of this study are available from the corresponding authors upon reasonable request.
\nocite{*}
\bibliography{HiRES_UED}

\providecommand{\noopsort}[1]{}\providecommand{\singleletter}[1]{#1}%
\begin{thebibliography}{73}%
\makeatletter
\providecommand \@ifxundefined [1]{%
 \@ifx{#1\undefined}
}%
\providecommand \@ifnum [1]{%
 \ifnum #1\expandafter \@firstoftwo
 \else \expandafter \@secondoftwo
 \fi
}%
\providecommand \@ifx [1]{%
 \ifx #1\expandafter \@firstoftwo
 \else \expandafter \@secondoftwo
 \fi
}%
\providecommand \natexlab [1]{#1}%
\providecommand \enquote  [1]{``#1''}%
\providecommand \bibnamefont  [1]{#1}%
\providecommand \bibfnamefont [1]{#1}%
\providecommand \citenamefont [1]{#1}%
\providecommand \href@noop [0]{\@secondoftwo}%
\providecommand \href [0]{\begingroup \@sanitize@url \@href}%
\providecommand \@href[1]{\@@startlink{#1}\@@href}%
\providecommand \@@href[1]{\endgroup#1\@@endlink}%
\providecommand \@sanitize@url [0]{\catcode `\\12\catcode `\$12\catcode
  `\&12\catcode `\#12\catcode `\^12\catcode `\_12\catcode `\%12\relax}%
\providecommand \@@startlink[1]{}%
\providecommand \@@endlink[0]{}%
\providecommand \url  [0]{\begingroup\@sanitize@url \@url }%
\providecommand \@url [1]{\endgroup\@href {#1}{\urlprefix }}%
\providecommand \urlprefix  [0]{URL }%
\providecommand \Eprint [0]{\href }%
\providecommand \doibase [0]{https://doi.org/}%
\providecommand \selectlanguage [0]{\@gobble}%
\providecommand \bibinfo  [0]{\@secondoftwo}%
\providecommand \bibfield  [0]{\@secondoftwo}%
\providecommand \translation [1]{[#1]}%
\providecommand \BibitemOpen [0]{}%
\providecommand \bibitemStop [0]{}%
\providecommand \bibitemNoStop [0]{.\EOS\space}%
\providecommand \EOS [0]{\spacefactor3000\relax}%
\providecommand \BibitemShut  [1]{\csname bibitem#1\endcsname}%
\let\auto@bib@innerbib\@empty
\bibitem [{\citenamefont {Mourou}\ and\ \citenamefont
  {Williamson}(1982)}]{Mourou1982}%
  \BibitemOpen
  \bibfield  {author} {\bibinfo {author} {\bibfnamefont {G.}~\bibnamefont
  {Mourou}}\ and\ \bibinfo {author} {\bibfnamefont {S.}~\bibnamefont
  {Williamson}},\ }\bibfield  {title} {\enquote {\bibinfo {title} {Picosecond
  electron diffraction},}\ }\href@noop {} {\bibfield  {journal} {\bibinfo
  {journal} {Applied Physics Letters}\ }\textbf {\bibinfo {volume} {41}},\
  \bibinfo {pages} {44--45} (\bibinfo {year} {1982})}\BibitemShut {NoStop}%
\bibitem [{\citenamefont {Ischenko}\ \emph {et~al.}(1983)\citenamefont
  {Ischenko}, \citenamefont {Golubkov}, \citenamefont {Spiridonov},
  \citenamefont {Zgurskii}, \citenamefont {Akhmanov}, \citenamefont
  {Vabischevich},\ and\ \citenamefont {Bagratashvili}}]{Ischenko1983}%
  \BibitemOpen
  \bibfield  {author} {\bibinfo {author} {\bibfnamefont {A.~A.}\ \bibnamefont
  {Ischenko}}, \bibinfo {author} {\bibfnamefont {V.~V.}\ \bibnamefont
  {Golubkov}}, \bibinfo {author} {\bibfnamefont {V.~P.}\ \bibnamefont
  {Spiridonov}}, \bibinfo {author} {\bibfnamefont {A.~V.}\ \bibnamefont
  {Zgurskii}}, \bibinfo {author} {\bibfnamefont {A.~S.}\ \bibnamefont
  {Akhmanov}}, \bibinfo {author} {\bibfnamefont {M.~G.}\ \bibnamefont
  {Vabischevich}},\ and\ \bibinfo {author} {\bibfnamefont {V.~N.}\ \bibnamefont
  {Bagratashvili}},\ }\bibfield  {title} {\enquote {\bibinfo {title} {A
  stroboscopical gas-electron diffraction method for the investigation of
  short-lived molecular species},}\ }\href@noop {} {\bibfield  {journal}
  {\bibinfo  {journal} {Applied Physics B}\ }\textbf {\bibinfo {volume} {32}},\
  \bibinfo {pages} {161--163} (\bibinfo {year} {1983})}\BibitemShut {NoStop}%
\bibitem [{\citenamefont {Filippetto}\ \emph {et~al.}(2022)\citenamefont
  {Filippetto}, \citenamefont {Musumeci}, \citenamefont {Li}, \citenamefont
  {Siwick}, \citenamefont {Otto}, \citenamefont {Centurion},\ and\
  \citenamefont {Nunes}}]{filippetto2022rmp}%
  \BibitemOpen
  \bibfield  {author} {\bibinfo {author} {\bibfnamefont {D.}~\bibnamefont
  {Filippetto}}, \bibinfo {author} {\bibfnamefont {P.}~\bibnamefont
  {Musumeci}}, \bibinfo {author} {\bibfnamefont {R.}~\bibnamefont {Li}},
  \bibinfo {author} {\bibfnamefont {B.~J.}\ \bibnamefont {Siwick}}, \bibinfo
  {author} {\bibfnamefont {M.}~\bibnamefont {Otto}}, \bibinfo {author}
  {\bibfnamefont {M.}~\bibnamefont {Centurion}},\ and\ \bibinfo {author}
  {\bibfnamefont {J.}~\bibnamefont {Nunes}},\ }\bibfield  {title} {\enquote
  {\bibinfo {title} {Ultrafast electron diffraction: Visualizing dynamic states
  of matter},}\ }\href@noop {} {\bibfield  {journal} {\bibinfo  {journal}
  {Reviews of Modern Physics}\ }\textbf {\bibinfo {volume} {94}},\ \bibinfo
  {pages} {045004} (\bibinfo {year} {2022})}\BibitemShut {NoStop}%
\bibitem [{\citenamefont {Kogar}\ \emph {et~al.}(2020)\citenamefont {Kogar},
  \citenamefont {Zong}, \citenamefont {Dolgirev}, \citenamefont {Shen},
  \citenamefont {Straquadine}, \citenamefont {Bie}, \citenamefont {Wang},
  \citenamefont {Rohwer}, \citenamefont {Tung}, \citenamefont {Yang},
  \citenamefont {Li}, \citenamefont {Yang}, \citenamefont {Weathersby},
  \citenamefont {Park}, \citenamefont {Kozina}, \citenamefont {Sie},
  \citenamefont {Wen}, \citenamefont {Jarillo-Herrero}, \citenamefont {Fisher},
  \citenamefont {Wang},\ and\ \citenamefont {Gedik}}]{Kogar2020}%
  \BibitemOpen
  \bibfield  {author} {\bibinfo {author} {\bibfnamefont {A.}~\bibnamefont
  {Kogar}}, \bibinfo {author} {\bibfnamefont {A.}~\bibnamefont {Zong}},
  \bibinfo {author} {\bibfnamefont {P.~E.}\ \bibnamefont {Dolgirev}}, \bibinfo
  {author} {\bibfnamefont {X.}~\bibnamefont {Shen}}, \bibinfo {author}
  {\bibfnamefont {J.}~\bibnamefont {Straquadine}}, \bibinfo {author}
  {\bibfnamefont {Y.-Q.}\ \bibnamefont {Bie}}, \bibinfo {author} {\bibfnamefont
  {X.}~\bibnamefont {Wang}}, \bibinfo {author} {\bibfnamefont {T.}~\bibnamefont
  {Rohwer}}, \bibinfo {author} {\bibfnamefont {I.-C.}\ \bibnamefont {Tung}},
  \bibinfo {author} {\bibfnamefont {Y.}~\bibnamefont {Yang}}, \bibinfo {author}
  {\bibfnamefont {R.}~\bibnamefont {Li}}, \bibinfo {author} {\bibfnamefont
  {J.}~\bibnamefont {Yang}}, \bibinfo {author} {\bibfnamefont {S.}~\bibnamefont
  {Weathersby}}, \bibinfo {author} {\bibfnamefont {S.}~\bibnamefont {Park}},
  \bibinfo {author} {\bibfnamefont {M.~E.}\ \bibnamefont {Kozina}}, \bibinfo
  {author} {\bibfnamefont {E.~J.}\ \bibnamefont {Sie}}, \bibinfo {author}
  {\bibfnamefont {H.}~\bibnamefont {Wen}}, \bibinfo {author} {\bibfnamefont
  {P.}~\bibnamefont {Jarillo-Herrero}}, \bibinfo {author} {\bibfnamefont
  {I.~R.}\ \bibnamefont {Fisher}}, \bibinfo {author} {\bibfnamefont
  {X.}~\bibnamefont {Wang}},\ and\ \bibinfo {author} {\bibfnamefont
  {N.}~\bibnamefont {Gedik}},\ }\bibfield  {title} {\enquote {\bibinfo {title}
  {Light-induced charge density wave in \ce{LaTe3}},}\ }\href@noop {}
  {\bibfield  {journal} {\bibinfo  {journal} {Nature Physics}\ }\textbf
  {\bibinfo {volume} {16}},\ \bibinfo {pages} {159--163} (\bibinfo {year}
  {2020})}\BibitemShut {NoStop}%
\bibitem [{\citenamefont {Gao}\ \emph {et~al.}(2013)\citenamefont {Gao},
  \citenamefont {Lu}, \citenamefont {Jean-Ruel}, \citenamefont {Liu},
  \citenamefont {Marx}, \citenamefont {Onda}, \citenamefont {Koshihara},
  \citenamefont {Nakano}, \citenamefont {Shao}, \citenamefont {Hiramatsu},
  \citenamefont {Saito}, \citenamefont {Yamochi}, \citenamefont {Cooney},
  \citenamefont {Moriena}, \citenamefont {Sciaini},\ and\ \citenamefont
  {Miller}}]{Gao2013}%
  \BibitemOpen
  \bibfield  {author} {\bibinfo {author} {\bibfnamefont {M.}~\bibnamefont
  {Gao}}, \bibinfo {author} {\bibfnamefont {C.}~\bibnamefont {Lu}}, \bibinfo
  {author} {\bibfnamefont {H.}~\bibnamefont {Jean-Ruel}}, \bibinfo {author}
  {\bibfnamefont {L.~C.}\ \bibnamefont {Liu}}, \bibinfo {author} {\bibfnamefont
  {A.}~\bibnamefont {Marx}}, \bibinfo {author} {\bibfnamefont {K.}~\bibnamefont
  {Onda}}, \bibinfo {author} {\bibfnamefont {S.-y.}\ \bibnamefont {Koshihara}},
  \bibinfo {author} {\bibfnamefont {Y.}~\bibnamefont {Nakano}}, \bibinfo
  {author} {\bibfnamefont {X.}~\bibnamefont {Shao}}, \bibinfo {author}
  {\bibfnamefont {T.}~\bibnamefont {Hiramatsu}}, \bibinfo {author}
  {\bibfnamefont {G.}~\bibnamefont {Saito}}, \bibinfo {author} {\bibfnamefont
  {H.}~\bibnamefont {Yamochi}}, \bibinfo {author} {\bibfnamefont {R.~R.}\
  \bibnamefont {Cooney}}, \bibinfo {author} {\bibfnamefont {G.}~\bibnamefont
  {Moriena}}, \bibinfo {author} {\bibfnamefont {G.}~\bibnamefont {Sciaini}},\
  and\ \bibinfo {author} {\bibfnamefont {R.~J.~D.}\ \bibnamefont {Miller}},\
  }\bibfield  {title} {\enquote {\bibinfo {title} {Mapping molecular motions
  leading to charge delocalization with ultrabright electrons},}\ }\href@noop
  {} {\bibfield  {journal} {\bibinfo  {journal} {Nature}\ }\textbf {\bibinfo
  {volume} {496}},\ \bibinfo {pages} {343--346} (\bibinfo {year}
  {2013})}\BibitemShut {NoStop}%
\bibitem [{\citenamefont {Dudek}\ and\ \citenamefont
  {Weber}(2001)}]{Dudek2001}%
  \BibitemOpen
  \bibfield  {author} {\bibinfo {author} {\bibfnamefont {R.~C.}\ \bibnamefont
  {Dudek}}\ and\ \bibinfo {author} {\bibfnamefont {P.~M.}\ \bibnamefont
  {Weber}},\ }\bibfield  {title} {\enquote {\bibinfo {title} {Ultrafast
  diffraction imaging of the electrocyclic ring-opening reaction of
  1,3-cyclohexadiene},}\ }\href@noop {} {\bibfield  {journal} {\bibinfo
  {journal} {J. Phys. Chem. A}\ }\textbf {\bibinfo {volume} {105}},\ \bibinfo
  {pages} {4167--4171} (\bibinfo {year} {2001})}\BibitemShut {NoStop}%
\bibitem [{\citenamefont {Cao}, \citenamefont {Ihee},\ and\ \citenamefont
  {Zewail}(1999)}]{Cao1999}%
  \BibitemOpen
  \bibfield  {author} {\bibinfo {author} {\bibfnamefont {J.}~\bibnamefont
  {Cao}}, \bibinfo {author} {\bibfnamefont {H.}~\bibnamefont {Ihee}},\ and\
  \bibinfo {author} {\bibfnamefont {A.~H.}\ \bibnamefont {Zewail}},\ }\bibfield
   {title} {\enquote {\bibinfo {title} {Ultrafast electron diffraction and
  direct observation of transient structures in a chemical reaction},}\
  }\href@noop {} {\bibfield  {journal} {\bibinfo  {journal} {Proceedings of the
  National Academy of Sciences}\ }\textbf {\bibinfo {volume} {96}},\ \bibinfo
  {pages} {338--342} (\bibinfo {year} {1999})}\BibitemShut {NoStop}%
\bibitem [{\citenamefont {Siwick}\ \emph {et~al.}(2003)\citenamefont {Siwick},
  \citenamefont {Dwyer}, \citenamefont {Jordan},\ and\ \citenamefont
  {Miller}}]{Siwick2003}%
  \BibitemOpen
  \bibfield  {author} {\bibinfo {author} {\bibfnamefont {B.~J.}\ \bibnamefont
  {Siwick}}, \bibinfo {author} {\bibfnamefont {J.~R.}\ \bibnamefont {Dwyer}},
  \bibinfo {author} {\bibfnamefont {R.~E.}\ \bibnamefont {Jordan}},\ and\
  \bibinfo {author} {\bibfnamefont {R.~J.~D.}\ \bibnamefont {Miller}},\
  }\bibfield  {title} {\enquote {\bibinfo {title} {An atomic-level view of
  melting using femtosecond electron diffraction},}\ }\href@noop {} {\bibfield
  {journal} {\bibinfo  {journal} {Science}\ }\textbf {\bibinfo {volume}
  {302}},\ \bibinfo {pages} {1382--1385} (\bibinfo {year} {2003})}\BibitemShut
  {NoStop}%
\bibitem [{\citenamefont {Wu}\ \emph {et~al.}(2022)\citenamefont {Wu},
  \citenamefont {Tang}, \citenamefont {Zhao}, \citenamefont {Zhu},
  \citenamefont {Jiang}, \citenamefont {Zou}, \citenamefont {Hong},
  \citenamefont {Luo}, \citenamefont {Xiang},\ and\ \citenamefont
  {Zhang}}]{Wu2022}%
  \BibitemOpen
  \bibfield  {author} {\bibinfo {author} {\bibfnamefont {J.}~\bibnamefont
  {Wu}}, \bibinfo {author} {\bibfnamefont {M.}~\bibnamefont {Tang}}, \bibinfo
  {author} {\bibfnamefont {L.}~\bibnamefont {Zhao}}, \bibinfo {author}
  {\bibfnamefont {P.}~\bibnamefont {Zhu}}, \bibinfo {author} {\bibfnamefont
  {T.}~\bibnamefont {Jiang}}, \bibinfo {author} {\bibfnamefont
  {X.}~\bibnamefont {Zou}}, \bibinfo {author} {\bibfnamefont {L.}~\bibnamefont
  {Hong}}, \bibinfo {author} {\bibfnamefont {S.-N.}\ \bibnamefont {Luo}},
  \bibinfo {author} {\bibfnamefont {D.}~\bibnamefont {Xiang}},\ and\ \bibinfo
  {author} {\bibfnamefont {J.}~\bibnamefont {Zhang}},\ }\bibfield  {title}
  {\enquote {\bibinfo {title} {Ultrafast atomic view of laser-induced melting
  and breathing motion of metallic liquid clusters with {MeV} ultrafast
  electron diffraction},}\ }\href@noop {} {\bibfield  {journal} {\bibinfo
  {journal} {Proceedings of the National Academy of Sciences}\ }\textbf
  {\bibinfo {volume} {119}},\ \bibinfo {pages} {e2111949119} (\bibinfo {year}
  {2022})}\BibitemShut {NoStop}%
\bibitem [{\citenamefont {Mo}\ \emph {et~al.}(2016{\natexlab{a}})\citenamefont
  {Mo}, \citenamefont {Shen}, \citenamefont {Chen}, \citenamefont {Li},
  \citenamefont {Dunning}, \citenamefont {Sokolowski-Tinten}, \citenamefont
  {Zheng}, \citenamefont {Weathersby}, \citenamefont {Reid}, \citenamefont
  {Coffee}, \citenamefont {Makasyuk}, \citenamefont {Edstrom}, \citenamefont
  {McCormick}, \citenamefont {Jobe}, \citenamefont {Hast}, \citenamefont
  {Glenzer},\ and\ \citenamefont {Wang}}]{Mo2016}%
  \BibitemOpen
  \bibfield  {author} {\bibinfo {author} {\bibfnamefont {M.~Z.}\ \bibnamefont
  {Mo}}, \bibinfo {author} {\bibfnamefont {X.}~\bibnamefont {Shen}}, \bibinfo
  {author} {\bibfnamefont {Z.}~\bibnamefont {Chen}}, \bibinfo {author}
  {\bibfnamefont {R.~K.}\ \bibnamefont {Li}}, \bibinfo {author} {\bibfnamefont
  {M.}~\bibnamefont {Dunning}}, \bibinfo {author} {\bibfnamefont
  {K.}~\bibnamefont {Sokolowski-Tinten}}, \bibinfo {author} {\bibfnamefont
  {Q.}~\bibnamefont {Zheng}}, \bibinfo {author} {\bibfnamefont {S.~P.}\
  \bibnamefont {Weathersby}}, \bibinfo {author} {\bibfnamefont {A.~H.}\
  \bibnamefont {Reid}}, \bibinfo {author} {\bibfnamefont {R.}~\bibnamefont
  {Coffee}}, \bibinfo {author} {\bibfnamefont {I.}~\bibnamefont {Makasyuk}},
  \bibinfo {author} {\bibfnamefont {S.}~\bibnamefont {Edstrom}}, \bibinfo
  {author} {\bibfnamefont {D.}~\bibnamefont {McCormick}}, \bibinfo {author}
  {\bibfnamefont {K.}~\bibnamefont {Jobe}}, \bibinfo {author} {\bibfnamefont
  {C.}~\bibnamefont {Hast}}, \bibinfo {author} {\bibfnamefont {S.~H.}\
  \bibnamefont {Glenzer}},\ and\ \bibinfo {author} {\bibfnamefont
  {X.}~\bibnamefont {Wang}},\ }\bibfield  {title} {\enquote {\bibinfo {title}
  {Single-shot mega-electronvolt ultrafast electron diffraction for structure
  dynamic studies of warm dense matter},}\ }\href@noop {} {\bibfield  {journal}
  {\bibinfo  {journal} {Review of Scientific Instruments}\ }\textbf {\bibinfo
  {volume} {87}},\ \bibinfo {pages} {11D810} (\bibinfo {year}
  {2016}{\natexlab{a}})}\BibitemShut {NoStop}%
\bibitem [{\citenamefont {Sun}\ \emph {et~al.}(2020)\citenamefont {Sun},
  \citenamefont {Sun}, \citenamefont {Bartles}, \citenamefont {Wozniak},
  \citenamefont {Williams}, \citenamefont {Zhang},\ and\ \citenamefont
  {Ruan}}]{Sun2020}%
  \BibitemOpen
  \bibfield  {author} {\bibinfo {author} {\bibfnamefont {S.}~\bibnamefont
  {Sun}}, \bibinfo {author} {\bibfnamefont {X.}~\bibnamefont {Sun}}, \bibinfo
  {author} {\bibfnamefont {D.}~\bibnamefont {Bartles}}, \bibinfo {author}
  {\bibfnamefont {E.}~\bibnamefont {Wozniak}}, \bibinfo {author} {\bibfnamefont
  {J.}~\bibnamefont {Williams}}, \bibinfo {author} {\bibfnamefont
  {P.}~\bibnamefont {Zhang}},\ and\ \bibinfo {author} {\bibfnamefont {C.-Y.}\
  \bibnamefont {Ruan}},\ }\bibfield  {title} {\enquote {\bibinfo {title}
  {Direct imaging of plasma waves using ultrafast electron microscopy},}\
  }\href@noop {} {\bibfield  {journal} {\bibinfo  {journal} {Structural
  Dynamics}\ }\textbf {\bibinfo {volume} {7}},\ \bibinfo {pages} {064301}
  (\bibinfo {year} {2020})}\BibitemShut {NoStop}%
\bibitem [{\citenamefont {Zandi}\ \emph {et~al.}(2020)\citenamefont {Zandi},
  \citenamefont {Sykes}, \citenamefont {Cornelius}, \citenamefont {Alcorn},
  \citenamefont {Zerbe}, \citenamefont {Duxbury}, \citenamefont {Reed},\ and\
  \citenamefont {van~der Veen}}]{Zandi2020}%
  \BibitemOpen
  \bibfield  {author} {\bibinfo {author} {\bibfnamefont {O.}~\bibnamefont
  {Zandi}}, \bibinfo {author} {\bibfnamefont {A.~E.}\ \bibnamefont {Sykes}},
  \bibinfo {author} {\bibfnamefont {R.~D.}\ \bibnamefont {Cornelius}}, \bibinfo
  {author} {\bibfnamefont {F.~M.}\ \bibnamefont {Alcorn}}, \bibinfo {author}
  {\bibfnamefont {B.~S.}\ \bibnamefont {Zerbe}}, \bibinfo {author}
  {\bibfnamefont {P.~M.}\ \bibnamefont {Duxbury}}, \bibinfo {author}
  {\bibfnamefont {B.~W.}\ \bibnamefont {Reed}},\ and\ \bibinfo {author}
  {\bibfnamefont {R.~M.}\ \bibnamefont {van~der Veen}},\ }\bibfield  {title}
  {\enquote {\bibinfo {title} {Transient lensing from a photoemitted electron
  gas imaged by ultrafast electron microscopy},}\ }\href@noop {} {\bibfield
  {journal} {\bibinfo  {journal} {Nature Communications}\ }\textbf {\bibinfo
  {volume} {11}},\ \bibinfo {pages} {3001} (\bibinfo {year}
  {2020})}\BibitemShut {NoStop}%
\bibitem [{\citenamefont {Jean-Ruel}\ \emph {et~al.}(2013)\citenamefont
  {Jean-Ruel}, \citenamefont {Gao}, \citenamefont {Kochman}, \citenamefont
  {Lu}, \citenamefont {Liu}, \citenamefont {Cooney}, \citenamefont {Morrison},\
  and\ \citenamefont {Miller}}]{Jean-Ruel2013}%
  \BibitemOpen
  \bibfield  {author} {\bibinfo {author} {\bibfnamefont {H.}~\bibnamefont
  {Jean-Ruel}}, \bibinfo {author} {\bibfnamefont {M.}~\bibnamefont {Gao}},
  \bibinfo {author} {\bibfnamefont {M.~A.}\ \bibnamefont {Kochman}}, \bibinfo
  {author} {\bibfnamefont {C.}~\bibnamefont {Lu}}, \bibinfo {author}
  {\bibfnamefont {L.~C.}\ \bibnamefont {Liu}}, \bibinfo {author} {\bibfnamefont
  {R.~R.}\ \bibnamefont {Cooney}}, \bibinfo {author} {\bibfnamefont {C.~A.}\
  \bibnamefont {Morrison}},\ and\ \bibinfo {author} {\bibfnamefont {R.~J.~D.}\
  \bibnamefont {Miller}},\ }\bibfield  {title} {\enquote {\bibinfo {title}
  {Ring-closing reaction in diarylethene captured by femtosecond electron
  crystallography},}\ }\href {https://doi.org/10.1021/jp409245h} {\bibfield
  {journal} {\bibinfo  {journal} {The Journal of Physical Chemistry B}\
  }\textbf {\bibinfo {volume} {117}},\ \bibinfo {pages} {15894--15902}
  (\bibinfo {year} {2013})}\BibitemShut {NoStop}%
\bibitem [{\citenamefont {Michalik}\ and\ \citenamefont
  {Sipe}(2006)}]{Michalik2006}%
  \BibitemOpen
  \bibfield  {author} {\bibinfo {author} {\bibfnamefont {A.~M.}\ \bibnamefont
  {Michalik}}\ and\ \bibinfo {author} {\bibfnamefont {J.~E.}\ \bibnamefont
  {Sipe}},\ }\bibfield  {title} {\enquote {\bibinfo {title} {Analytic model of
  electron pulse propagation in ultrafast electron diffraction experiments},}\
  }\href@noop {} {\bibfield  {journal} {\bibinfo  {journal} {Journal of Applied
  Physics}\ }\textbf {\bibinfo {volume} {99}},\ \bibinfo {pages} {054908}
  (\bibinfo {year} {2006})}\BibitemShut {NoStop}%
\bibitem [{\citenamefont {Reed}(2006)}]{Reed2006}%
  \BibitemOpen
  \bibfield  {author} {\bibinfo {author} {\bibfnamefont {B.~W.}\ \bibnamefont
  {Reed}},\ }\bibfield  {title} {\enquote {\bibinfo {title} {Femtosecond
  electron pulse propagation for ultrafast electron diffraction},}\ }\href@noop
  {} {\bibfield  {journal} {\bibinfo  {journal} {Journal of Applied Physics}\
  }\textbf {\bibinfo {volume} {100}},\ \bibinfo {pages} {034916} (\bibinfo
  {year} {2006})}\BibitemShut {NoStop}%
\bibitem [{\citenamefont {Storeck}, \citenamefont {Rossnagel},\ and\
  \citenamefont {Ropers}(2021)}]{Storeck2021}%
  \BibitemOpen
  \bibfield  {author} {\bibinfo {author} {\bibfnamefont {G.}~\bibnamefont
  {Storeck}}, \bibinfo {author} {\bibfnamefont {K.}~\bibnamefont {Rossnagel}},\
  and\ \bibinfo {author} {\bibfnamefont {C.}~\bibnamefont {Ropers}},\
  }\bibfield  {title} {\enquote {\bibinfo {title} {Ultrafast spot-profile
  {LEED} of a charge-density wave phase transition},}\ }\href@noop {}
  {\bibfield  {journal} {\bibinfo  {journal} {Applied Physics Letters}\
  }\textbf {\bibinfo {volume} {118}},\ \bibinfo {pages} {221603} (\bibinfo
  {year} {2021})}\BibitemShut {NoStop}%
\bibitem [{\citenamefont {Barwick}\ \emph {et~al.}(2008)\citenamefont
  {Barwick}, \citenamefont {Park}, \citenamefont {Kwon}, \citenamefont
  {Baskin},\ and\ \citenamefont {Zewail}}]{Barwick2008}%
  \BibitemOpen
  \bibfield  {author} {\bibinfo {author} {\bibfnamefont {B.}~\bibnamefont
  {Barwick}}, \bibinfo {author} {\bibfnamefont {H.~S.}\ \bibnamefont {Park}},
  \bibinfo {author} {\bibfnamefont {O.-H.}\ \bibnamefont {Kwon}}, \bibinfo
  {author} {\bibfnamefont {J.~S.}\ \bibnamefont {Baskin}},\ and\ \bibinfo
  {author} {\bibfnamefont {A.~H.}\ \bibnamefont {Zewail}},\ }\bibfield  {title}
  {\enquote {\bibinfo {title} {{4D} imaging of transient structures and
  morphologies in ultrafast electron microscopy},}\ }\href@noop {} {\bibfield
  {journal} {\bibinfo  {journal} {Science}\ }\textbf {\bibinfo {volume}
  {322}},\ \bibinfo {pages} {1227--1231} (\bibinfo {year} {2008})}\BibitemShut
  {NoStop}%
\bibitem [{\citenamefont {Aidelsburger}\ \emph {et~al.}(2010)\citenamefont
  {Aidelsburger}, \citenamefont {Kirchner}, \citenamefont {Krausz},\ and\
  \citenamefont {Baum}}]{Aidelsburger2010}%
  \BibitemOpen
  \bibfield  {author} {\bibinfo {author} {\bibfnamefont {M.}~\bibnamefont
  {Aidelsburger}}, \bibinfo {author} {\bibfnamefont {F.~O.}\ \bibnamefont
  {Kirchner}}, \bibinfo {author} {\bibfnamefont {F.}~\bibnamefont {Krausz}},\
  and\ \bibinfo {author} {\bibfnamefont {P.}~\bibnamefont {Baum}},\ }\bibfield
  {title} {\enquote {\bibinfo {title} {Single-electron pulses for ultrafast
  diffraction},}\ }\href@noop {} {\bibfield  {journal} {\bibinfo  {journal}
  {Proceedings of the National Academy of Sciences}\ }\textbf {\bibinfo
  {volume} {107}},\ \bibinfo {pages} {19714--19719} (\bibinfo {year}
  {2010})}\BibitemShut {NoStop}%
\bibitem [{\citenamefont {LaGrange}\ \emph {et~al.}(2006)\citenamefont
  {LaGrange}, \citenamefont {Armstrong}, \citenamefont {Boyden}, \citenamefont
  {Brown}, \citenamefont {Campbell}, \citenamefont {Colvin}, \citenamefont
  {DeHope}, \citenamefont {Frank}, \citenamefont {Gibson}, \citenamefont
  {Hartemann}, \citenamefont {Kim}, \citenamefont {King}, \citenamefont {Pyke},
  \citenamefont {Reed}, \citenamefont {Shirk}, \citenamefont {Shuttlesworth},
  \citenamefont {Stuart}, \citenamefont {Torralva},\ and\ \citenamefont
  {Browning}}]{LaGrange2006}%
  \BibitemOpen
  \bibfield  {author} {\bibinfo {author} {\bibfnamefont {T.}~\bibnamefont
  {LaGrange}}, \bibinfo {author} {\bibfnamefont {M.~R.}\ \bibnamefont
  {Armstrong}}, \bibinfo {author} {\bibfnamefont {K.}~\bibnamefont {Boyden}},
  \bibinfo {author} {\bibfnamefont {C.~G.}\ \bibnamefont {Brown}}, \bibinfo
  {author} {\bibfnamefont {G.~H.}\ \bibnamefont {Campbell}}, \bibinfo {author}
  {\bibfnamefont {J.~D.}\ \bibnamefont {Colvin}}, \bibinfo {author}
  {\bibfnamefont {W.~J.}\ \bibnamefont {DeHope}}, \bibinfo {author}
  {\bibfnamefont {A.~M.}\ \bibnamefont {Frank}}, \bibinfo {author}
  {\bibfnamefont {D.~J.}\ \bibnamefont {Gibson}}, \bibinfo {author}
  {\bibfnamefont {F.~V.}\ \bibnamefont {Hartemann}}, \bibinfo {author}
  {\bibfnamefont {J.~S.}\ \bibnamefont {Kim}}, \bibinfo {author} {\bibfnamefont
  {W.~E.}\ \bibnamefont {King}}, \bibinfo {author} {\bibfnamefont {B.~J.}\
  \bibnamefont {Pyke}}, \bibinfo {author} {\bibfnamefont {B.~W.}\ \bibnamefont
  {Reed}}, \bibinfo {author} {\bibfnamefont {M.~D.}\ \bibnamefont {Shirk}},
  \bibinfo {author} {\bibfnamefont {R.~M.}\ \bibnamefont {Shuttlesworth}},
  \bibinfo {author} {\bibfnamefont {B.~C.}\ \bibnamefont {Stuart}}, \bibinfo
  {author} {\bibfnamefont {B.~R.}\ \bibnamefont {Torralva}},\ and\ \bibinfo
  {author} {\bibfnamefont {N.~D.}\ \bibnamefont {Browning}},\ }\bibfield
  {title} {\enquote {\bibinfo {title} {Single-shot dynamic transmission
  electron microscopy},}\ }\href@noop {} {\bibfield  {journal} {\bibinfo
  {journal} {Applied Physics Letters}\ }\textbf {\bibinfo {volume} {89}},\
  \bibinfo {pages} {044105} (\bibinfo {year} {2006})}\BibitemShut {NoStop}%
\bibitem [{\citenamefont {Chatelain}\ \emph {et~al.}(2012)\citenamefont
  {Chatelain}, \citenamefont {Morrison}, \citenamefont {Godbout},\ and\
  \citenamefont {Siwick}}]{Chatelain2012}%
  \BibitemOpen
  \bibfield  {author} {\bibinfo {author} {\bibfnamefont {R.~P.}\ \bibnamefont
  {Chatelain}}, \bibinfo {author} {\bibfnamefont {V.~R.}\ \bibnamefont
  {Morrison}}, \bibinfo {author} {\bibfnamefont {C.}~\bibnamefont {Godbout}},\
  and\ \bibinfo {author} {\bibfnamefont {B.~J.}\ \bibnamefont {Siwick}},\
  }\bibfield  {title} {\enquote {\bibinfo {title} {Ultrafast electron
  diffraction with radio-frequency compressed electron pulses},}\ }\href@noop
  {} {\bibfield  {journal} {\bibinfo  {journal} {Applied Physics Letters}\
  }\textbf {\bibinfo {volume} {101}},\ \bibinfo {pages} {081901} (\bibinfo
  {year} {2012})}\BibitemShut {NoStop}%
\bibitem [{\citenamefont {Zhang}\ \emph {et~al.}(2021)\citenamefont {Zhang},
  \citenamefont {Kroh}, \citenamefont {Ritzkowsky}, \citenamefont {Rohwer},
  \citenamefont {Fakhari}, \citenamefont {Cankaya}, \citenamefont {Calendron},
  \citenamefont {Matlis},\ and\ \citenamefont {Kärtner}}]{Zhang2021}%
  \BibitemOpen
  \bibfield  {author} {\bibinfo {author} {\bibfnamefont {D.}~\bibnamefont
  {Zhang}}, \bibinfo {author} {\bibfnamefont {T.}~\bibnamefont {Kroh}},
  \bibinfo {author} {\bibfnamefont {F.}~\bibnamefont {Ritzkowsky}}, \bibinfo
  {author} {\bibfnamefont {T.}~\bibnamefont {Rohwer}}, \bibinfo {author}
  {\bibfnamefont {M.}~\bibnamefont {Fakhari}}, \bibinfo {author} {\bibfnamefont
  {H.}~\bibnamefont {Cankaya}}, \bibinfo {author} {\bibfnamefont {A.-L.}\
  \bibnamefont {Calendron}}, \bibinfo {author} {\bibfnamefont {N.~H.}\
  \bibnamefont {Matlis}},\ and\ \bibinfo {author} {\bibfnamefont {F.~X.}\
  \bibnamefont {Kärtner}},\ }\bibfield  {title} {\enquote {\bibinfo {title}
  {{THz}-enhanced {DC} ultrafast electron diffractometer},}\ }\href@noop {}
  {\bibfield  {journal} {\bibinfo  {journal} {Ultrafast Science}\ }\textbf
  {\bibinfo {volume} {2021}} (\bibinfo {year} {2021})}\BibitemShut {NoStop}%
\bibitem [{\citenamefont {Musumeci}, \citenamefont {Moody},\ and\ \citenamefont
  {Scoby}(2008)}]{Musumeci2008}%
  \BibitemOpen
  \bibfield  {author} {\bibinfo {author} {\bibfnamefont {P.}~\bibnamefont
  {Musumeci}}, \bibinfo {author} {\bibfnamefont {J.}~\bibnamefont {Moody}},\
  and\ \bibinfo {author} {\bibfnamefont {C.}~\bibnamefont {Scoby}},\ }\bibfield
   {title} {\enquote {\bibinfo {title} {Relativistic electron diffraction at
  the {UCLA} {Pegasus} photoinjector laboratory},}\ }\href@noop {} {\bibfield
  {journal} {\bibinfo  {journal} {Ultramicroscopy}\ }\textbf {\bibinfo {volume}
  {108}},\ \bibinfo {pages} {1450--1453} (\bibinfo {year} {2008})}\BibitemShut
  {NoStop}%
\bibitem [{\citenamefont {Hastings}\ \emph {et~al.}(2006)\citenamefont
  {Hastings}, \citenamefont {Rudakov}, \citenamefont {Dowell}, \citenamefont
  {Schmerge}, \citenamefont {Cardoza}, \citenamefont {Castro}, \citenamefont
  {Gierman}, \citenamefont {Loos},\ and\ \citenamefont {Weber}}]{SLAC_first}%
  \BibitemOpen
  \bibfield  {author} {\bibinfo {author} {\bibfnamefont {J.~B.}\ \bibnamefont
  {Hastings}}, \bibinfo {author} {\bibfnamefont {F.~M.}\ \bibnamefont
  {Rudakov}}, \bibinfo {author} {\bibfnamefont {D.~H.}\ \bibnamefont {Dowell}},
  \bibinfo {author} {\bibfnamefont {J.~F.}\ \bibnamefont {Schmerge}}, \bibinfo
  {author} {\bibfnamefont {J.~D.}\ \bibnamefont {Cardoza}}, \bibinfo {author}
  {\bibfnamefont {J.~M.}\ \bibnamefont {Castro}}, \bibinfo {author}
  {\bibfnamefont {S.~M.}\ \bibnamefont {Gierman}}, \bibinfo {author}
  {\bibfnamefont {H.}~\bibnamefont {Loos}},\ and\ \bibinfo {author}
  {\bibfnamefont {P.~M.}\ \bibnamefont {Weber}},\ }\bibfield  {title} {\enquote
  {\bibinfo {title} {Ultrafast time-resolved electron diffraction with megavolt
  electron beams},}\ }\href@noop {} {\bibfield  {journal} {\bibinfo  {journal}
  {Applied Physics Letters}\ }\textbf {\bibinfo {volume} {89}},\ \bibinfo
  {pages} {184109} (\bibinfo {year} {2006})}\BibitemShut {NoStop}%
\bibitem [{\citenamefont {Ren\'e~de Cotret}\ \emph {et~al.}(2019)\citenamefont
  {Ren\'e~de Cotret}, \citenamefont {P\"ohls}, \citenamefont {Stern},
  \citenamefont {Otto}, \citenamefont {Sutton},\ and\ \citenamefont
  {Siwick}}]{Cotret2019}%
  \BibitemOpen
  \bibfield  {author} {\bibinfo {author} {\bibfnamefont {L.~P.}\ \bibnamefont
  {Ren\'e~de Cotret}}, \bibinfo {author} {\bibfnamefont {J.-H.}\ \bibnamefont
  {P\"ohls}}, \bibinfo {author} {\bibfnamefont {M.~J.}\ \bibnamefont {Stern}},
  \bibinfo {author} {\bibfnamefont {M.~R.}\ \bibnamefont {Otto}}, \bibinfo
  {author} {\bibfnamefont {M.}~\bibnamefont {Sutton}},\ and\ \bibinfo {author}
  {\bibfnamefont {B.~J.}\ \bibnamefont {Siwick}},\ }\bibfield  {title}
  {\enquote {\bibinfo {title} {Time- and momentum-resolved phonon population
  dynamics with ultrafast electron diffuse scattering},}\ }\href@noop {}
  {\bibfield  {journal} {\bibinfo  {journal} {Phys. Rev. B}\ }\textbf {\bibinfo
  {volume} {100}},\ \bibinfo {pages} {214115} (\bibinfo {year}
  {2019})}\BibitemShut {NoStop}%
\bibitem [{\citenamefont {Zuo}\ \emph {et~al.}(2004)\citenamefont {Zuo},
  \citenamefont {Gao}, \citenamefont {Tao}, \citenamefont {Li}, \citenamefont
  {Twesten},\ and\ \citenamefont {Petrov}}]{Zuo2004}%
  \BibitemOpen
  \bibfield  {author} {\bibinfo {author} {\bibfnamefont {J.}~\bibnamefont
  {Zuo}}, \bibinfo {author} {\bibfnamefont {M.}~\bibnamefont {Gao}}, \bibinfo
  {author} {\bibfnamefont {J.}~\bibnamefont {Tao}}, \bibinfo {author}
  {\bibfnamefont {B.}~\bibnamefont {Li}}, \bibinfo {author} {\bibfnamefont
  {R.}~\bibnamefont {Twesten}},\ and\ \bibinfo {author} {\bibfnamefont
  {I.}~\bibnamefont {Petrov}},\ }\bibfield  {title} {\enquote {\bibinfo {title}
  {Coherent nano-area electron diffraction},}\ }\href
  {https://doi.org/https://doi.org/10.1002/jemt.20096} {\bibfield  {journal}
  {\bibinfo  {journal} {Microscopy Research and Technique}\ }\textbf {\bibinfo
  {volume} {64}},\ \bibinfo {pages} {347--355} (\bibinfo {year}
  {2004})}\BibitemShut {NoStop}%
\bibitem [{\citenamefont {Carlino}\ \emph {et~al.}(2018)\citenamefont
  {Carlino}, \citenamefont {Scattarella}, \citenamefont {Caro}, \citenamefont
  {Giannini}, \citenamefont {Siliqi}, \citenamefont {Colombo},\ and\
  \citenamefont {Galli}}]{Elvio2018}%
  \BibitemOpen
  \bibfield  {author} {\bibinfo {author} {\bibfnamefont {E.}~\bibnamefont
  {Carlino}}, \bibinfo {author} {\bibfnamefont {F.}~\bibnamefont
  {Scattarella}}, \bibinfo {author} {\bibfnamefont {L.~D.}\ \bibnamefont
  {Caro}}, \bibinfo {author} {\bibfnamefont {C.}~\bibnamefont {Giannini}},
  \bibinfo {author} {\bibfnamefont {D.}~\bibnamefont {Siliqi}}, \bibinfo
  {author} {\bibfnamefont {A.}~\bibnamefont {Colombo}},\ and\ \bibinfo {author}
  {\bibfnamefont {D.~E.}\ \bibnamefont {Galli}},\ }\bibfield  {title} {\enquote
  {\bibinfo {title} {Coherent diffraction imaging in transmission electron
  microscopy for atomic resolution quantitative studies of the matter},}\
  }\href@noop {} {\bibfield  {journal} {\bibinfo  {journal} {Materials}\
  }\textbf {\bibinfo {volume} {11}} (\bibinfo {year} {2018})}\BibitemShut
  {NoStop}%
\bibitem [{\citenamefont {Sannibale}\ \emph {et~al.}(2012)\citenamefont
  {Sannibale}, \citenamefont {Filippetto}, \citenamefont {Papadopoulos},
  \citenamefont {Staples}, \citenamefont {Wells}, \citenamefont {Bailey},
  \citenamefont {Baptiste}, \citenamefont {Corlett}, \citenamefont {Cork},
  \citenamefont {De~Santis}, \citenamefont {Dimaggio}, \citenamefont
  {Doolittle}, \citenamefont {Doyle}, \citenamefont {Feng}, \citenamefont
  {Garcia~Quintas}, \citenamefont {Huang}, \citenamefont {Huang}, \citenamefont
  {Kramasz}, \citenamefont {Kwiatkowski}, \citenamefont {Lellinger},
  \citenamefont {Moroz}, \citenamefont {Norum}, \citenamefont {Padmore},
  \citenamefont {Pappas}, \citenamefont {Portmann}, \citenamefont {Vecchione},
  \citenamefont {Vinco}, \citenamefont {Zolotorev},\ and\ \citenamefont
  {Zucca}}]{PhysRevSTAB.15.103501}%
  \BibitemOpen
  \bibfield  {author} {\bibinfo {author} {\bibfnamefont {F.}~\bibnamefont
  {Sannibale}}, \bibinfo {author} {\bibfnamefont {D.}~\bibnamefont
  {Filippetto}}, \bibinfo {author} {\bibfnamefont {C.~F.}\ \bibnamefont
  {Papadopoulos}}, \bibinfo {author} {\bibfnamefont {J.}~\bibnamefont
  {Staples}}, \bibinfo {author} {\bibfnamefont {R.}~\bibnamefont {Wells}},
  \bibinfo {author} {\bibfnamefont {B.}~\bibnamefont {Bailey}}, \bibinfo
  {author} {\bibfnamefont {K.}~\bibnamefont {Baptiste}}, \bibinfo {author}
  {\bibfnamefont {J.}~\bibnamefont {Corlett}}, \bibinfo {author} {\bibfnamefont
  {C.}~\bibnamefont {Cork}}, \bibinfo {author} {\bibfnamefont {S.}~\bibnamefont
  {De~Santis}}, \bibinfo {author} {\bibfnamefont {S.}~\bibnamefont {Dimaggio}},
  \bibinfo {author} {\bibfnamefont {L.}~\bibnamefont {Doolittle}}, \bibinfo
  {author} {\bibfnamefont {J.}~\bibnamefont {Doyle}}, \bibinfo {author}
  {\bibfnamefont {J.}~\bibnamefont {Feng}}, \bibinfo {author} {\bibfnamefont
  {D.}~\bibnamefont {Garcia~Quintas}}, \bibinfo {author} {\bibfnamefont
  {G.}~\bibnamefont {Huang}}, \bibinfo {author} {\bibfnamefont
  {H.}~\bibnamefont {Huang}}, \bibinfo {author} {\bibfnamefont
  {T.}~\bibnamefont {Kramasz}}, \bibinfo {author} {\bibfnamefont
  {S.}~\bibnamefont {Kwiatkowski}}, \bibinfo {author} {\bibfnamefont
  {R.}~\bibnamefont {Lellinger}}, \bibinfo {author} {\bibfnamefont
  {V.}~\bibnamefont {Moroz}}, \bibinfo {author} {\bibfnamefont {W.~E.}\
  \bibnamefont {Norum}}, \bibinfo {author} {\bibfnamefont {H.}~\bibnamefont
  {Padmore}}, \bibinfo {author} {\bibfnamefont {C.}~\bibnamefont {Pappas}},
  \bibinfo {author} {\bibfnamefont {G.}~\bibnamefont {Portmann}}, \bibinfo
  {author} {\bibfnamefont {T.}~\bibnamefont {Vecchione}}, \bibinfo {author}
  {\bibfnamefont {M.}~\bibnamefont {Vinco}}, \bibinfo {author} {\bibfnamefont
  {M.}~\bibnamefont {Zolotorev}},\ and\ \bibinfo {author} {\bibfnamefont
  {F.}~\bibnamefont {Zucca}},\ }\bibfield  {title} {\enquote {\bibinfo {title}
  {Advanced photoinjector experiment photogun commissioning results},}\
  }\href@noop {} {\bibfield  {journal} {\bibinfo  {journal} {Phys. Rev. ST
  Accel. Beams}\ }\textbf {\bibinfo {volume} {15}},\ \bibinfo {pages} {103501}
  (\bibinfo {year} {2012})}\BibitemShut {NoStop}%
\bibitem [{\citenamefont {Bazarov}\ \emph {et~al.}(2011)\citenamefont
  {Bazarov}, \citenamefont {Cultrera}, \citenamefont {Bartnik}, \citenamefont
  {Dunham}, \citenamefont {Karkare}, \citenamefont {Li}, \citenamefont {Liu},
  \citenamefont {Maxson},\ and\ \citenamefont {Roussel}}]{10.1063/1.3596450}%
  \BibitemOpen
  \bibfield  {author} {\bibinfo {author} {\bibfnamefont {I.}~\bibnamefont
  {Bazarov}}, \bibinfo {author} {\bibfnamefont {L.}~\bibnamefont {Cultrera}},
  \bibinfo {author} {\bibfnamefont {A.}~\bibnamefont {Bartnik}}, \bibinfo
  {author} {\bibfnamefont {B.}~\bibnamefont {Dunham}}, \bibinfo {author}
  {\bibfnamefont {S.}~\bibnamefont {Karkare}}, \bibinfo {author} {\bibfnamefont
  {Y.}~\bibnamefont {Li}}, \bibinfo {author} {\bibfnamefont {X.}~\bibnamefont
  {Liu}}, \bibinfo {author} {\bibfnamefont {J.}~\bibnamefont {Maxson}},\ and\
  \bibinfo {author} {\bibfnamefont {W.}~\bibnamefont {Roussel}},\ }\bibfield
  {title} {\enquote {\bibinfo {title} {{Thermal emittance measurements of a
  cesium potassium antimonide photocathode}},}\ }\href
  {https://doi.org/10.1063/1.3596450} {\bibfield  {journal} {\bibinfo
  {journal} {Applied Physics Letters}\ }\textbf {\bibinfo {volume} {98}}
  (\bibinfo {year} {2011}),\ 10.1063/1.3596450},\ \bibinfo {note} {224101},\
  \Eprint
  {https://arxiv.org/abs/https://pubs.aip.org/aip/apl/article-pdf/doi/10.1063/1.3596450/13353170/224101\_1\_online.pdf}
  {https://pubs.aip.org/aip/apl/article-pdf/doi/10.1063/1.3596450/13353170/224101\_1\_online.pdf}
  \BibitemShut {NoStop}%
\bibitem [{GPT()}]{GPT}%
  \BibitemOpen
  \bibfield  {title} {\enquote {\bibinfo {title} {General particle tracer},}\
  }\href@noop {} {\bibinfo  {journal} {http://www.pulsar.nl/gpt/}\
  }\BibitemShut {NoStop}%
\bibitem [{\citenamefont {Cropp}\ \emph {et~al.}(2023)\citenamefont {Cropp},
  \citenamefont {Moos}, \citenamefont {Scheinker}, \citenamefont {Gilardi},
  \citenamefont {Wang}, \citenamefont {Paiagua}, \citenamefont {Serrano},
  \citenamefont {Musumeci},\ and\ \citenamefont
  {Filippetto}}]{cropp2023virtual}%
  \BibitemOpen
\bibfield  {journal} {  }\bibfield  {author} {\bibinfo {author} {\bibfnamefont
  {F.}~\bibnamefont {Cropp}}, \bibinfo {author} {\bibfnamefont
  {L.}~\bibnamefont {Moos}}, \bibinfo {author} {\bibfnamefont {A.}~\bibnamefont
  {Scheinker}}, \bibinfo {author} {\bibfnamefont {A.}~\bibnamefont {Gilardi}},
  \bibinfo {author} {\bibfnamefont {D.}~\bibnamefont {Wang}}, \bibinfo {author}
  {\bibfnamefont {S.}~\bibnamefont {Paiagua}}, \bibinfo {author} {\bibfnamefont
  {C.}~\bibnamefont {Serrano}}, \bibinfo {author} {\bibfnamefont
  {P.}~\bibnamefont {Musumeci}},\ and\ \bibinfo {author} {\bibfnamefont
  {D.}~\bibnamefont {Filippetto}},\ }\bibfield  {title} {\enquote {\bibinfo
  {title} {Virtual-diagnostic-based time stamping for ultrafast electron
  diffraction},}\ }\href@noop {} {\bibfield  {journal} {\bibinfo  {journal}
  {arXiv preprint arXiv:2302.04916}\ } (\bibinfo {year} {2023})}\BibitemShut
  {NoStop}%
\bibitem [{\citenamefont {Filippetto}\ and\ \citenamefont
  {Qian}(2016)}]{Filippetto_2016}%
  \BibitemOpen
  \bibfield  {author} {\bibinfo {author} {\bibfnamefont {D.}~\bibnamefont
  {Filippetto}}\ and\ \bibinfo {author} {\bibfnamefont {H.}~\bibnamefont
  {Qian}},\ }\bibfield  {title} {\enquote {\bibinfo {title} {Design of a
  high-flux instrument for ultrafast electron diffraction and microscopy},}\
  }\href@noop {} {\bibfield  {journal} {\bibinfo  {journal} {Journal of Physics
  B: Atomic, Molecular and Optical Physics}\ }\textbf {\bibinfo {volume}
  {49}},\ \bibinfo {pages} {104003} (\bibinfo {year} {2016})}\BibitemShut
  {NoStop}%
\bibitem [{\citenamefont {Siddiqui}\ \emph {et~al.}(2021)\citenamefont
  {Siddiqui}, \citenamefont {Durham}, \citenamefont {Cropp}, \citenamefont
  {Ophus}, \citenamefont {Rajpurohit}, \citenamefont {Zhu}, \citenamefont
  {Carlström}, \citenamefont {Stavrakas}, \citenamefont {Mao}, \citenamefont
  {Raja}, \citenamefont {Musumeci}, \citenamefont {Tan}, \citenamefont {Minor},
  \citenamefont {Filippetto},\ and\ \citenamefont {Kaindl}}]{Siddiqui2021}%
  \BibitemOpen
  \bibfield  {author} {\bibinfo {author} {\bibfnamefont {K.~M.}\ \bibnamefont
  {Siddiqui}}, \bibinfo {author} {\bibfnamefont {D.~B.}\ \bibnamefont
  {Durham}}, \bibinfo {author} {\bibfnamefont {F.}~\bibnamefont {Cropp}},
  \bibinfo {author} {\bibfnamefont {C.}~\bibnamefont {Ophus}}, \bibinfo
  {author} {\bibfnamefont {S.}~\bibnamefont {Rajpurohit}}, \bibinfo {author}
  {\bibfnamefont {Y.}~\bibnamefont {Zhu}}, \bibinfo {author} {\bibfnamefont
  {J.~D.}\ \bibnamefont {Carlström}}, \bibinfo {author} {\bibfnamefont
  {C.}~\bibnamefont {Stavrakas}}, \bibinfo {author} {\bibfnamefont
  {Z.}~\bibnamefont {Mao}}, \bibinfo {author} {\bibfnamefont {A.}~\bibnamefont
  {Raja}}, \bibinfo {author} {\bibfnamefont {P.}~\bibnamefont {Musumeci}},
  \bibinfo {author} {\bibfnamefont {L.~Z.}\ \bibnamefont {Tan}}, \bibinfo
  {author} {\bibfnamefont {A.~M.}\ \bibnamefont {Minor}}, \bibinfo {author}
  {\bibfnamefont {D.}~\bibnamefont {Filippetto}},\ and\ \bibinfo {author}
  {\bibfnamefont {R.~A.}\ \bibnamefont {Kaindl}},\ }\bibfield  {title}
  {\enquote {\bibinfo {title} {Ultrafast optical melting of trimer
  superstructure in layered \ce{1T}'-\ce{TaTe2}},}\ }\href@noop {} {\bibfield
  {journal} {\bibinfo  {journal} {Communications Physics}\ }\textbf {\bibinfo
  {volume} {4}},\ \bibinfo {pages} {152} (\bibinfo {year} {2021})}\BibitemShut
  {NoStop}%
\bibitem [{\citenamefont {Durham}\ \emph {et~al.}(2022)\citenamefont {Durham},
  \citenamefont {Ophus}, \citenamefont {Siddiqui}, \citenamefont {Minor},\ and\
  \citenamefont {Filippetto}}]{Durham2022}%
  \BibitemOpen
  \bibfield  {author} {\bibinfo {author} {\bibfnamefont {D.~B.}\ \bibnamefont
  {Durham}}, \bibinfo {author} {\bibfnamefont {C.}~\bibnamefont {Ophus}},
  \bibinfo {author} {\bibfnamefont {K.~M.}\ \bibnamefont {Siddiqui}}, \bibinfo
  {author} {\bibfnamefont {A.~M.}\ \bibnamefont {Minor}},\ and\ \bibinfo
  {author} {\bibfnamefont {D.}~\bibnamefont {Filippetto}},\ }\bibfield  {title}
  {\enquote {\bibinfo {title} {Accurate quantification of lattice temperature
  dynamics from ultrafast electron diffraction of single-crystal films using
  dynamical scattering simulations},}\ }\href@noop {} {\bibfield  {journal}
  {\bibinfo  {journal} {Structural Dynamics}\ }\textbf {\bibinfo {volume}
  {9}},\ \bibinfo {pages} {064302} (\bibinfo {year} {2022})}\BibitemShut
  {NoStop}%
\bibitem [{\citenamefont {Wang}\ \emph {et~al.}(2022)\citenamefont {Wang},
  \citenamefont {Li}, \citenamefont {Wu}, \citenamefont {Sears}, \citenamefont
  {Ji}, \citenamefont {Shen}, \citenamefont {Reid}, \citenamefont {Tao},
  \citenamefont {Robinson}, \citenamefont {Zhu},\ and\ \citenamefont
  {Dean}}]{Wang2022}%
  \BibitemOpen
  \bibfield  {author} {\bibinfo {author} {\bibfnamefont {W.}~\bibnamefont
  {Wang}}, \bibinfo {author} {\bibfnamefont {J.}~\bibnamefont {Li}}, \bibinfo
  {author} {\bibfnamefont {L.}~\bibnamefont {Wu}}, \bibinfo {author}
  {\bibfnamefont {J.}~\bibnamefont {Sears}}, \bibinfo {author} {\bibfnamefont
  {F.}~\bibnamefont {Ji}}, \bibinfo {author} {\bibfnamefont {X.}~\bibnamefont
  {Shen}}, \bibinfo {author} {\bibfnamefont {A.~H.}\ \bibnamefont {Reid}},
  \bibinfo {author} {\bibfnamefont {J.}~\bibnamefont {Tao}}, \bibinfo {author}
  {\bibfnamefont {I.~K.}\ \bibnamefont {Robinson}}, \bibinfo {author}
  {\bibfnamefont {Y.}~\bibnamefont {Zhu}},\ and\ \bibinfo {author}
  {\bibfnamefont {M.~P.~M.}\ \bibnamefont {Dean}},\ }\bibfield  {title}
  {\enquote {\bibinfo {title} {Dual-stage structural response to quenching
  charge order in magnetite},}\ }\href@noop {} {\bibfield  {journal} {\bibinfo
  {journal} {Phys. Rev. B}\ }\textbf {\bibinfo {volume} {106}},\ \bibinfo
  {pages} {195131} (\bibinfo {year} {2022})}\BibitemShut {NoStop}%
\bibitem [{\citenamefont {Li}\ \emph {et~al.}(2022)\citenamefont {Li},
  \citenamefont {Wu}, \citenamefont {Yang}, \citenamefont {Jin}, \citenamefont
  {Wang}, \citenamefont {Tao}, \citenamefont {Boatner}, \citenamefont
  {Babzien}, \citenamefont {Fedurin}, \citenamefont {Palmer}, \citenamefont
  {Yin}, \citenamefont {Delaire},\ and\ \citenamefont {Zhu}}]{JLi2022}%
  \BibitemOpen
  \bibfield  {author} {\bibinfo {author} {\bibfnamefont {J.}~\bibnamefont
  {Li}}, \bibinfo {author} {\bibfnamefont {L.}~\bibnamefont {Wu}}, \bibinfo
  {author} {\bibfnamefont {S.}~\bibnamefont {Yang}}, \bibinfo {author}
  {\bibfnamefont {X.}~\bibnamefont {Jin}}, \bibinfo {author} {\bibfnamefont
  {W.}~\bibnamefont {Wang}}, \bibinfo {author} {\bibfnamefont {J.}~\bibnamefont
  {Tao}}, \bibinfo {author} {\bibfnamefont {L.}~\bibnamefont {Boatner}},
  \bibinfo {author} {\bibfnamefont {M.}~\bibnamefont {Babzien}}, \bibinfo
  {author} {\bibfnamefont {M.}~\bibnamefont {Fedurin}}, \bibinfo {author}
  {\bibfnamefont {M.}~\bibnamefont {Palmer}}, \bibinfo {author} {\bibfnamefont
  {W.}~\bibnamefont {Yin}}, \bibinfo {author} {\bibfnamefont {O.}~\bibnamefont
  {Delaire}},\ and\ \bibinfo {author} {\bibfnamefont {Y.}~\bibnamefont {Zhu}},\
  }\bibfield  {title} {\enquote {\bibinfo {title} {Direct detection of v-v atom
  dimerization and rotation dynamic pathways upon ultrafast photoexcitation in
  ${\mathrm{vo}}_{2}$},}\ }\href@noop {} {\bibfield  {journal} {\bibinfo
  {journal} {Phys. Rev. X}\ }\textbf {\bibinfo {volume} {12}},\ \bibinfo
  {pages} {021032} (\bibinfo {year} {2022})}\BibitemShut {NoStop}%
\bibitem [{\citenamefont {Sciaini}\ \emph {et~al.}(2009)\citenamefont
  {Sciaini}, \citenamefont {Harb}, \citenamefont {Kruglik}, \citenamefont
  {Payer}, \citenamefont {Hebeisen}, \citenamefont {Heringdorf}, \citenamefont
  {Yamaguchi}, \citenamefont {Hoegen}, \citenamefont {Ernstorfer},\ and\
  \citenamefont {Miller}}]{Sciaini2009}%
  \BibitemOpen
  \bibfield  {author} {\bibinfo {author} {\bibfnamefont {G.}~\bibnamefont
  {Sciaini}}, \bibinfo {author} {\bibfnamefont {M.}~\bibnamefont {Harb}},
  \bibinfo {author} {\bibfnamefont {S.~G.}\ \bibnamefont {Kruglik}}, \bibinfo
  {author} {\bibfnamefont {T.}~\bibnamefont {Payer}}, \bibinfo {author}
  {\bibfnamefont {C.~T.}\ \bibnamefont {Hebeisen}}, \bibinfo {author}
  {\bibfnamefont {F.-J. M.~z.}\ \bibnamefont {Heringdorf}}, \bibinfo {author}
  {\bibfnamefont {M.}~\bibnamefont {Yamaguchi}}, \bibinfo {author}
  {\bibfnamefont {M.~H.-v.}\ \bibnamefont {Hoegen}}, \bibinfo {author}
  {\bibfnamefont {R.}~\bibnamefont {Ernstorfer}},\ and\ \bibinfo {author}
  {\bibfnamefont {R.~J.~D.}\ \bibnamefont {Miller}},\ }\bibfield  {title}
  {\enquote {\bibinfo {title} {Electronic acceleration of atomic motions and
  disordering in bismuth},}\ }\href@noop {} {\bibfield  {journal} {\bibinfo
  {journal} {Nature}\ }\textbf {\bibinfo {volume} {458}},\ \bibinfo {pages}
  {56--59} (\bibinfo {year} {2009})}\BibitemShut {NoStop}%
\bibitem [{\citenamefont {Qi}\ \emph {et~al.}(2022)\citenamefont {Qi},
  \citenamefont {Chen}, \citenamefont {Vasileiadis}, \citenamefont {Zahn},
  \citenamefont {Seiler}, \citenamefont {Li},\ and\ \citenamefont
  {Ernstorfer}}]{Qi2022}%
  \BibitemOpen
  \bibfield  {author} {\bibinfo {author} {\bibfnamefont {Y.}~\bibnamefont
  {Qi}}, \bibinfo {author} {\bibfnamefont {N.}~\bibnamefont {Chen}}, \bibinfo
  {author} {\bibfnamefont {T.}~\bibnamefont {Vasileiadis}}, \bibinfo {author}
  {\bibfnamefont {D.}~\bibnamefont {Zahn}}, \bibinfo {author} {\bibfnamefont
  {H.}~\bibnamefont {Seiler}}, \bibinfo {author} {\bibfnamefont
  {X.}~\bibnamefont {Li}},\ and\ \bibinfo {author} {\bibfnamefont
  {R.}~\bibnamefont {Ernstorfer}},\ }\bibfield  {title} {\enquote {\bibinfo
  {title} {Photoinduced ultrafast transition of the local correlated structure
  in chalcogenide phase-change materials},}\ }\href@noop {} {\bibfield
  {journal} {\bibinfo  {journal} {Phys. Rev. Lett.}\ }\textbf {\bibinfo
  {volume} {129}},\ \bibinfo {pages} {135701} (\bibinfo {year}
  {2022})}\BibitemShut {NoStop}%
\bibitem [{\citenamefont {Tauchert}\ \emph {et~al.}(2022)\citenamefont
  {Tauchert}, \citenamefont {Volkov}, \citenamefont {Ehberger}, \citenamefont
  {Kazenwadel}, \citenamefont {Evers}, \citenamefont {Lange}, \citenamefont
  {Donges}, \citenamefont {Book}, \citenamefont {Kreuzpaintner}, \citenamefont
  {Nowak},\ and\ \citenamefont {Baum}}]{Tauchert2022}%
  \BibitemOpen
  \bibfield  {author} {\bibinfo {author} {\bibfnamefont {S.~R.}\ \bibnamefont
  {Tauchert}}, \bibinfo {author} {\bibfnamefont {M.}~\bibnamefont {Volkov}},
  \bibinfo {author} {\bibfnamefont {D.}~\bibnamefont {Ehberger}}, \bibinfo
  {author} {\bibfnamefont {D.}~\bibnamefont {Kazenwadel}}, \bibinfo {author}
  {\bibfnamefont {M.}~\bibnamefont {Evers}}, \bibinfo {author} {\bibfnamefont
  {H.}~\bibnamefont {Lange}}, \bibinfo {author} {\bibfnamefont
  {A.}~\bibnamefont {Donges}}, \bibinfo {author} {\bibfnamefont
  {A.}~\bibnamefont {Book}}, \bibinfo {author} {\bibfnamefont {W.}~\bibnamefont
  {Kreuzpaintner}}, \bibinfo {author} {\bibfnamefont {U.}~\bibnamefont
  {Nowak}},\ and\ \bibinfo {author} {\bibfnamefont {P.}~\bibnamefont {Baum}},\
  }\bibfield  {title} {\enquote {\bibinfo {title} {Polarized phonons carry
  angular momentum in ultrafast demagnetization},}\ }\href@noop {} {\bibfield
  {journal} {\bibinfo  {journal} {Nature}\ }\textbf {\bibinfo {volume} {602}},\
  \bibinfo {pages} {73--77} (\bibinfo {year} {2022})}\BibitemShut {NoStop}%
\bibitem [{\citenamefont {Sood}\ \emph {et~al.}(2023)\citenamefont {Sood},
  \citenamefont {Haber}, \citenamefont {Carlström}, \citenamefont {Peterson},
  \citenamefont {Barre}, \citenamefont {Georgaras}, \citenamefont {Reid},
  \citenamefont {Shen}, \citenamefont {Zajac}, \citenamefont {Regan},
  \citenamefont {Yang}, \citenamefont {Taniguchi}, \citenamefont {Watanabe},
  \citenamefont {Wang}, \citenamefont {Wang}, \citenamefont {Neaton},
  \citenamefont {Heinz}, \citenamefont {Lindenberg}, \citenamefont
  {da~Jornada},\ and\ \citenamefont {Raja}}]{Sood2023}%
  \BibitemOpen
  \bibfield  {author} {\bibinfo {author} {\bibfnamefont {A.}~\bibnamefont
  {Sood}}, \bibinfo {author} {\bibfnamefont {J.~B.}\ \bibnamefont {Haber}},
  \bibinfo {author} {\bibfnamefont {J.}~\bibnamefont {Carlström}}, \bibinfo
  {author} {\bibfnamefont {E.~A.}\ \bibnamefont {Peterson}}, \bibinfo {author}
  {\bibfnamefont {E.}~\bibnamefont {Barre}}, \bibinfo {author} {\bibfnamefont
  {J.~D.}\ \bibnamefont {Georgaras}}, \bibinfo {author} {\bibfnamefont
  {A.~H.~M.}\ \bibnamefont {Reid}}, \bibinfo {author} {\bibfnamefont
  {X.}~\bibnamefont {Shen}}, \bibinfo {author} {\bibfnamefont {M.~E.}\
  \bibnamefont {Zajac}}, \bibinfo {author} {\bibfnamefont {E.~C.}\ \bibnamefont
  {Regan}}, \bibinfo {author} {\bibfnamefont {J.}~\bibnamefont {Yang}},
  \bibinfo {author} {\bibfnamefont {T.}~\bibnamefont {Taniguchi}}, \bibinfo
  {author} {\bibfnamefont {K.}~\bibnamefont {Watanabe}}, \bibinfo {author}
  {\bibfnamefont {F.}~\bibnamefont {Wang}}, \bibinfo {author} {\bibfnamefont
  {X.}~\bibnamefont {Wang}}, \bibinfo {author} {\bibfnamefont {J.~B.}\
  \bibnamefont {Neaton}}, \bibinfo {author} {\bibfnamefont {T.~F.}\
  \bibnamefont {Heinz}}, \bibinfo {author} {\bibfnamefont {A.~M.}\ \bibnamefont
  {Lindenberg}}, \bibinfo {author} {\bibfnamefont {F.~H.}\ \bibnamefont
  {da~Jornada}},\ and\ \bibinfo {author} {\bibfnamefont {A.}~\bibnamefont
  {Raja}},\ }\bibfield  {title} {\enquote {\bibinfo {title} {Bidirectional
  phonon emission in two-dimensional heterostructures triggered by ultrafast
  charge transfer},}\ }\href@noop {} {\bibfield  {journal} {\bibinfo  {journal}
  {Nature Nanotechnology}\ }\textbf {\bibinfo {volume} {18}},\ \bibinfo {pages}
  {29--35} (\bibinfo {year} {2023})}\BibitemShut {NoStop}%
\bibitem [{\citenamefont {Gulde}\ \emph {et~al.}(2014)\citenamefont {Gulde},
  \citenamefont {Schweda}, \citenamefont {Storeck}, \citenamefont {Maiti},
  \citenamefont {Yu}, \citenamefont {Wodtke}, \citenamefont {Schäfer},\ and\
  \citenamefont {Ropers}}]{Gulde_2014}%
  \BibitemOpen
  \bibfield  {author} {\bibinfo {author} {\bibfnamefont {M.}~\bibnamefont
  {Gulde}}, \bibinfo {author} {\bibfnamefont {S.}~\bibnamefont {Schweda}},
  \bibinfo {author} {\bibfnamefont {G.}~\bibnamefont {Storeck}}, \bibinfo
  {author} {\bibfnamefont {M.}~\bibnamefont {Maiti}}, \bibinfo {author}
  {\bibfnamefont {H.~K.}\ \bibnamefont {Yu}}, \bibinfo {author} {\bibfnamefont
  {A.~M.}\ \bibnamefont {Wodtke}}, \bibinfo {author} {\bibfnamefont
  {S.}~\bibnamefont {Schäfer}},\ and\ \bibinfo {author} {\bibfnamefont
  {C.}~\bibnamefont {Ropers}},\ }\bibfield  {title} {\enquote {\bibinfo {title}
  {Ultrafast low-energy electron diffraction in transmission resolves
  polymer/graphene superstructure dynamics},}\ }\href@noop {} {\bibfield
  {journal} {\bibinfo  {journal} {Science}\ }\textbf {\bibinfo {volume}
  {345}},\ \bibinfo {pages} {200--204} (\bibinfo {year} {2014})}\BibitemShut
  {NoStop}%
\bibitem [{\citenamefont {Danz}, \citenamefont {Domr{\"o}se},\ and\
  \citenamefont {Ropers}(2021)}]{danz2021ultrafast}%
  \BibitemOpen
  \bibfield  {author} {\bibinfo {author} {\bibfnamefont {T.}~\bibnamefont
  {Danz}}, \bibinfo {author} {\bibfnamefont {T.}~\bibnamefont {Domr{\"o}se}},\
  and\ \bibinfo {author} {\bibfnamefont {C.}~\bibnamefont {Ropers}},\
  }\bibfield  {title} {\enquote {\bibinfo {title} {Ultrafast nanoimaging of the
  order parameter in a structural phase transition},}\ }\href@noop {}
  {\bibfield  {journal} {\bibinfo  {journal} {Science}\ }\textbf {\bibinfo
  {volume} {371}},\ \bibinfo {pages} {371--374} (\bibinfo {year}
  {2021})}\BibitemShut {NoStop}%
\bibitem [{\citenamefont {Otto}\ \emph {et~al.}(2019)\citenamefont {Otto},
  \citenamefont {de~Cotret}, \citenamefont {Valverde-Chavez}, \citenamefont
  {Tiwari}, \citenamefont {{\'E}mond}, \citenamefont {Chaker}, \citenamefont
  {Cooke},\ and\ \citenamefont {Siwick}}]{otto2019vo2}%
  \BibitemOpen
  \bibfield  {author} {\bibinfo {author} {\bibfnamefont {M.~R.}\ \bibnamefont
  {Otto}}, \bibinfo {author} {\bibfnamefont {L.~P.~R.}\ \bibnamefont
  {de~Cotret}}, \bibinfo {author} {\bibfnamefont {D.~A.}\ \bibnamefont
  {Valverde-Chavez}}, \bibinfo {author} {\bibfnamefont {K.~L.}\ \bibnamefont
  {Tiwari}}, \bibinfo {author} {\bibfnamefont {N.}~\bibnamefont {{\'E}mond}},
  \bibinfo {author} {\bibfnamefont {M.}~\bibnamefont {Chaker}}, \bibinfo
  {author} {\bibfnamefont {D.~G.}\ \bibnamefont {Cooke}},\ and\ \bibinfo
  {author} {\bibfnamefont {B.~J.}\ \bibnamefont {Siwick}},\ }\bibfield  {title}
  {\enquote {\bibinfo {title} {How optical excitation controls the structure
  and properties of vanadium dioxide},}\ }\href@noop {} {\bibfield  {journal}
  {\bibinfo  {journal} {Proceedings of the National Academy of Sciences}\
  }\textbf {\bibinfo {volume} {116}},\ \bibinfo {pages} {450--455} (\bibinfo
  {year} {2019})}\BibitemShut {NoStop}%
\bibitem [{\citenamefont {Zong}\ \emph {et~al.}(2019)\citenamefont {Zong},
  \citenamefont {Kogar}, \citenamefont {Bie}, \citenamefont {Rohwer},
  \citenamefont {Lee}, \citenamefont {Baldini}, \citenamefont {Erge{\c{c}}en},
  \citenamefont {Yilmaz}, \citenamefont {Freelon}, \citenamefont {Sie} \emph
  {et~al.}}]{gedik2019defects}%
  \BibitemOpen
  \bibfield  {author} {\bibinfo {author} {\bibfnamefont {A.}~\bibnamefont
  {Zong}}, \bibinfo {author} {\bibfnamefont {A.}~\bibnamefont {Kogar}},
  \bibinfo {author} {\bibfnamefont {Y.-Q.}\ \bibnamefont {Bie}}, \bibinfo
  {author} {\bibfnamefont {T.}~\bibnamefont {Rohwer}}, \bibinfo {author}
  {\bibfnamefont {C.}~\bibnamefont {Lee}}, \bibinfo {author} {\bibfnamefont
  {E.}~\bibnamefont {Baldini}}, \bibinfo {author} {\bibfnamefont
  {E.}~\bibnamefont {Erge{\c{c}}en}}, \bibinfo {author} {\bibfnamefont {M.~B.}\
  \bibnamefont {Yilmaz}}, \bibinfo {author} {\bibfnamefont {B.}~\bibnamefont
  {Freelon}}, \bibinfo {author} {\bibfnamefont {E.~J.}\ \bibnamefont {Sie}},
  \emph {et~al.},\ }\bibfield  {title} {\enquote {\bibinfo {title} {Evidence
  for topological defects in a photoinduced phase transition},}\ }\href@noop {}
  {\bibfield  {journal} {\bibinfo  {journal} {Nature Physics}\ }\textbf
  {\bibinfo {volume} {15}},\ \bibinfo {pages} {27--31} (\bibinfo {year}
  {2019})}\BibitemShut {NoStop}%
\bibitem [{\citenamefont {Johnson}\ \emph {et~al.}(2023)\citenamefont
  {Johnson}, \citenamefont {Perez-Salinas}, \citenamefont {Siddiqui},
  \citenamefont {Kim}, \citenamefont {Choi}, \citenamefont {Volckaert},
  \citenamefont {Majchrzak}, \citenamefont {Ulstrup}, \citenamefont {Agarwal},
  \citenamefont {Hallman}, \citenamefont {Haglund}, \citenamefont {Günther},
  \citenamefont {Pfau}, \citenamefont {Eisebitt}, \citenamefont {Backes},
  \citenamefont {Maccherozzi}, \citenamefont {Fitzpatrick}, \citenamefont
  {Dhesi}, \citenamefont {Gargiani}, \citenamefont {Valvidares}, \citenamefont
  {Artrith}, \citenamefont {de~Groot}, \citenamefont {Choi}, \citenamefont
  {Jang}, \citenamefont {Katoch}, \citenamefont {Kwon}, \citenamefont {Park},
  \citenamefont {Kim},\ and\ \citenamefont {Wall}}]{Johnson2023}%
  \BibitemOpen
  \bibfield  {author} {\bibinfo {author} {\bibfnamefont {A.~S.}\ \bibnamefont
  {Johnson}}, \bibinfo {author} {\bibfnamefont {D.}~\bibnamefont
  {Perez-Salinas}}, \bibinfo {author} {\bibfnamefont {K.~M.}\ \bibnamefont
  {Siddiqui}}, \bibinfo {author} {\bibfnamefont {S.}~\bibnamefont {Kim}},
  \bibinfo {author} {\bibfnamefont {S.}~\bibnamefont {Choi}}, \bibinfo {author}
  {\bibfnamefont {K.}~\bibnamefont {Volckaert}}, \bibinfo {author}
  {\bibfnamefont {P.~E.}\ \bibnamefont {Majchrzak}}, \bibinfo {author}
  {\bibfnamefont {S.}~\bibnamefont {Ulstrup}}, \bibinfo {author} {\bibfnamefont
  {N.}~\bibnamefont {Agarwal}}, \bibinfo {author} {\bibfnamefont
  {K.}~\bibnamefont {Hallman}}, \bibinfo {author} {\bibfnamefont {R.~F.}\
  \bibnamefont {Haglund}}, \bibinfo {author} {\bibfnamefont {C.~M.}\
  \bibnamefont {Günther}}, \bibinfo {author} {\bibfnamefont {B.}~\bibnamefont
  {Pfau}}, \bibinfo {author} {\bibfnamefont {S.}~\bibnamefont {Eisebitt}},
  \bibinfo {author} {\bibfnamefont {D.}~\bibnamefont {Backes}}, \bibinfo
  {author} {\bibfnamefont {F.}~\bibnamefont {Maccherozzi}}, \bibinfo {author}
  {\bibfnamefont {A.}~\bibnamefont {Fitzpatrick}}, \bibinfo {author}
  {\bibfnamefont {S.~S.}\ \bibnamefont {Dhesi}}, \bibinfo {author}
  {\bibfnamefont {P.}~\bibnamefont {Gargiani}}, \bibinfo {author}
  {\bibfnamefont {M.}~\bibnamefont {Valvidares}}, \bibinfo {author}
  {\bibfnamefont {N.}~\bibnamefont {Artrith}}, \bibinfo {author} {\bibfnamefont
  {F.}~\bibnamefont {de~Groot}}, \bibinfo {author} {\bibfnamefont
  {H.}~\bibnamefont {Choi}}, \bibinfo {author} {\bibfnamefont {D.}~\bibnamefont
  {Jang}}, \bibinfo {author} {\bibfnamefont {A.}~\bibnamefont {Katoch}},
  \bibinfo {author} {\bibfnamefont {S.}~\bibnamefont {Kwon}}, \bibinfo {author}
  {\bibfnamefont {S.~H.}\ \bibnamefont {Park}}, \bibinfo {author}
  {\bibfnamefont {H.}~\bibnamefont {Kim}},\ and\ \bibinfo {author}
  {\bibfnamefont {S.~E.}\ \bibnamefont {Wall}},\ }\bibfield  {title} {\enquote
  {\bibinfo {title} {Ultrafast x-ray imaging of the light-induced phase
  transition in vo2},}\ }\href@noop {} {\bibfield  {journal} {\bibinfo
  {journal} {Nature Physics}\ }\textbf {\bibinfo {volume} {19}},\ \bibinfo
  {pages} {215--220} (\bibinfo {year} {2023})}\BibitemShut {NoStop}%
\bibitem [{\citenamefont {Ji}\ \emph {et~al.}(2019{\natexlab{a}})\citenamefont
  {Ji}, \citenamefont {Durham}, \citenamefont {Minor}, \citenamefont
  {Musumeci}, \citenamefont {Navarro},\ and\ \citenamefont
  {Filippetto}}]{ji2019ultrafast}%
  \BibitemOpen
  \bibfield  {author} {\bibinfo {author} {\bibfnamefont {F.}~\bibnamefont
  {Ji}}, \bibinfo {author} {\bibfnamefont {D.~B.}\ \bibnamefont {Durham}},
  \bibinfo {author} {\bibfnamefont {A.~M.}\ \bibnamefont {Minor}}, \bibinfo
  {author} {\bibfnamefont {P.}~\bibnamefont {Musumeci}}, \bibinfo {author}
  {\bibfnamefont {J.~G.}\ \bibnamefont {Navarro}},\ and\ \bibinfo {author}
  {\bibfnamefont {D.}~\bibnamefont {Filippetto}},\ }\bibfield  {title}
  {\enquote {\bibinfo {title} {Ultrafast relativistic electron nanoprobes},}\
  }\href@noop {} {\bibfield  {journal} {\bibinfo  {journal} {Communications
  Physics}\ }\textbf {\bibinfo {volume} {2}},\ \bibinfo {pages} {1--10}
  (\bibinfo {year} {2019}{\natexlab{a}})}\BibitemShut {NoStop}%
\bibitem [{\citenamefont {Ji}\ \emph {et~al.}(2019{\natexlab{b}})\citenamefont
  {Ji}, \citenamefont {Navarro}, \citenamefont {Musumeci}, \citenamefont
  {Durham}, \citenamefont {Minor},\ and\ \citenamefont
  {Filippetto}}]{ji2019knife}%
  \BibitemOpen
  \bibfield  {author} {\bibinfo {author} {\bibfnamefont {F.}~\bibnamefont
  {Ji}}, \bibinfo {author} {\bibfnamefont {J.~G.}\ \bibnamefont {Navarro}},
  \bibinfo {author} {\bibfnamefont {P.}~\bibnamefont {Musumeci}}, \bibinfo
  {author} {\bibfnamefont {D.~B.}\ \bibnamefont {Durham}}, \bibinfo {author}
  {\bibfnamefont {A.~M.}\ \bibnamefont {Minor}},\ and\ \bibinfo {author}
  {\bibfnamefont {D.}~\bibnamefont {Filippetto}},\ }\bibfield  {title}
  {\enquote {\bibinfo {title} {Knife-edge based measurement of the {4D}
  transverse phase space of electron beams with picometer-scale emittance},}\
  }\href@noop {} {\bibfield  {journal} {\bibinfo  {journal} {Physical Review
  Accelerators and Beams}\ }\textbf {\bibinfo {volume} {22}},\ \bibinfo {pages}
  {082801} (\bibinfo {year} {2019}{\natexlab{b}})}\BibitemShut {NoStop}%
\bibitem [{\citenamefont {Zewail}(1994)}]{ZewailBook94}%
  \BibitemOpen
  \bibfield  {author} {\bibinfo {author} {\bibfnamefont {A.~H.}\ \bibnamefont
  {Zewail}},\ }\href@noop {} {\emph {\bibinfo {title} {Femtochemistry:
  Ultrafast Dynamics of the Chemical Bond}}}\ (\bibinfo  {publisher} {World
  Scientific Publishing Company},\ \bibinfo {year} {1994})\BibitemShut
  {NoStop}%
\bibitem [{\citenamefont {Warren}, \citenamefont {Rabitz},\ and\ \citenamefont
  {Dahleh}(1993)}]{Warren1993}%
  \BibitemOpen
  \bibfield  {author} {\bibinfo {author} {\bibfnamefont {W.~S.}\ \bibnamefont
  {Warren}}, \bibinfo {author} {\bibfnamefont {H.}~\bibnamefont {Rabitz}},\
  and\ \bibinfo {author} {\bibfnamefont {M.}~\bibnamefont {Dahleh}},\
  }\bibfield  {title} {\enquote {\bibinfo {title} {Coherent control of quantum
  dynamics: The dream is alive},}\ }\href@noop {} {\bibfield  {journal}
  {\bibinfo  {journal} {Science}\ }\textbf {\bibinfo {volume} {259}},\ \bibinfo
  {pages} {1581--1589} (\bibinfo {year} {1993})}\BibitemShut {NoStop}%
\bibitem [{\citenamefont {Deb}\ and\ \citenamefont {Weber}(2011)}]{Deb2011}%
  \BibitemOpen
  \bibfield  {author} {\bibinfo {author} {\bibfnamefont {S.}~\bibnamefont
  {Deb}}\ and\ \bibinfo {author} {\bibfnamefont {P.~M.}\ \bibnamefont
  {Weber}},\ }\bibfield  {title} {\enquote {\bibinfo {title} {The ultrafast
  pathway of photon-induced electrocyclic ring-opening reactions: The case of
  1,3-cyclohexadiene},}\ }\href@noop {} {\bibfield  {journal} {\bibinfo
  {journal} {Annual Review of Physical Chemistry}\ }\textbf {\bibinfo {volume}
  {62}},\ \bibinfo {pages} {19--39} (\bibinfo {year} {2011})}\BibitemShut
  {NoStop}%
\bibitem [{\citenamefont {Shen}\ \emph {et~al.}(2019)\citenamefont {Shen},
  \citenamefont {Nunes}, \citenamefont {Yang}, \citenamefont {Jobe},
  \citenamefont {Li}, \citenamefont {Lin}, \citenamefont {Moore}, \citenamefont
  {Niebuhr}, \citenamefont {Weathersby}, \citenamefont {Wolf}, \citenamefont
  {Yoneda}, \citenamefont {Guehr}, \citenamefont {Centurion},\ and\
  \citenamefont {Wang}}]{Shen2019}%
  \BibitemOpen
  \bibfield  {author} {\bibinfo {author} {\bibfnamefont {X.}~\bibnamefont
  {Shen}}, \bibinfo {author} {\bibfnamefont {J.~P.~F.}\ \bibnamefont {Nunes}},
  \bibinfo {author} {\bibfnamefont {J.}~\bibnamefont {Yang}}, \bibinfo {author}
  {\bibfnamefont {R.~K.}\ \bibnamefont {Jobe}}, \bibinfo {author}
  {\bibfnamefont {R.~K.}\ \bibnamefont {Li}}, \bibinfo {author} {\bibfnamefont
  {M.-F.}\ \bibnamefont {Lin}}, \bibinfo {author} {\bibfnamefont
  {B.}~\bibnamefont {Moore}}, \bibinfo {author} {\bibfnamefont
  {M.}~\bibnamefont {Niebuhr}}, \bibinfo {author} {\bibfnamefont {S.~P.}\
  \bibnamefont {Weathersby}}, \bibinfo {author} {\bibfnamefont {T.~J.~A.}\
  \bibnamefont {Wolf}}, \bibinfo {author} {\bibfnamefont {C.}~\bibnamefont
  {Yoneda}}, \bibinfo {author} {\bibfnamefont {M.}~\bibnamefont {Guehr}},
  \bibinfo {author} {\bibfnamefont {M.}~\bibnamefont {Centurion}},\ and\
  \bibinfo {author} {\bibfnamefont {X.~J.}\ \bibnamefont {Wang}},\ }\bibfield
  {title} {\enquote {\bibinfo {title} {Femtosecond gas-phase mega-electron-volt
  ultrafast electron diffraction},}\ }\href@noop {} {\bibfield  {journal}
  {\bibinfo  {journal} {Structural Dynamics}\ }\textbf {\bibinfo {volume}
  {6}},\ \bibinfo {pages} {054305} (\bibinfo {year} {2019})}\BibitemShut
  {NoStop}%
\bibitem [{\citenamefont {Ischenko}, \citenamefont {Weber},\ and\ \citenamefont
  {Miller}(2017)}]{MillerAndIschenko2017}%
  \BibitemOpen
  \bibfield  {author} {\bibinfo {author} {\bibfnamefont {A.~A.}\ \bibnamefont
  {Ischenko}}, \bibinfo {author} {\bibfnamefont {P.~M.}\ \bibnamefont
  {Weber}},\ and\ \bibinfo {author} {\bibfnamefont {R.~J.~D.}\ \bibnamefont
  {Miller}},\ }\bibfield  {title} {\enquote {\bibinfo {title} {Capturing
  chemistry in action with electrons: Realization of atomically resolved
  reaction dynamics},}\ }\href@noop {} {\bibfield  {journal} {\bibinfo
  {journal} {Chem. Rev.}\ }\textbf {\bibinfo {volume} {117}},\ \bibinfo {pages}
  {11066--11124} (\bibinfo {year} {2017})}\BibitemShut {NoStop}%
\bibitem [{\citenamefont {Ihee}\ \emph {et~al.}(2001)\citenamefont {Ihee},
  \citenamefont {Lobastov}, \citenamefont {Gomez}, \citenamefont {Goodson},
  \citenamefont {Srinivasan}, \citenamefont {Ruan},\ and\ \citenamefont
  {Zewail}}]{Ihee2001}%
  \BibitemOpen
  \bibfield  {author} {\bibinfo {author} {\bibfnamefont {H.}~\bibnamefont
  {Ihee}}, \bibinfo {author} {\bibfnamefont {V.~A.}\ \bibnamefont {Lobastov}},
  \bibinfo {author} {\bibfnamefont {U.~M.}\ \bibnamefont {Gomez}}, \bibinfo
  {author} {\bibfnamefont {B.~M.}\ \bibnamefont {Goodson}}, \bibinfo {author}
  {\bibfnamefont {R.}~\bibnamefont {Srinivasan}}, \bibinfo {author}
  {\bibfnamefont {C.-Y.}\ \bibnamefont {Ruan}},\ and\ \bibinfo {author}
  {\bibfnamefont {A.~H.}\ \bibnamefont {Zewail}},\ }\bibfield  {title}
  {\enquote {\bibinfo {title} {Direct imaging of transient molecular structures
  with ultrafast diffraction},}\ }\href@noop {} {\bibfield  {journal} {\bibinfo
   {journal} {Science}\ }\textbf {\bibinfo {volume} {291}},\ \bibinfo {pages}
  {458--462} (\bibinfo {year} {2001})}\BibitemShut {NoStop}%
\bibitem [{\citenamefont {Yang}\ \emph {et~al.}(2018)\citenamefont {Yang},
  \citenamefont {Zhu}, \citenamefont {Wolf}, \citenamefont {Li}, \citenamefont
  {Nunes}, \citenamefont {Coffee}, \citenamefont {Cryan}, \citenamefont
  {Gühr}, \citenamefont {Hegazy}, \citenamefont {Heinz}, \citenamefont {Jobe},
  \citenamefont {Li}, \citenamefont {Shen}, \citenamefont {Veccione},
  \citenamefont {Weathersby}, \citenamefont {Wilkin}, \citenamefont {Yoneda},
  \citenamefont {Zheng}, \citenamefont {Martinez}, \citenamefont {Centurion},\
  and\ \citenamefont {Wang}}]{Yang2018}%
  \BibitemOpen
  \bibfield  {author} {\bibinfo {author} {\bibfnamefont {J.}~\bibnamefont
  {Yang}}, \bibinfo {author} {\bibfnamefont {X.}~\bibnamefont {Zhu}}, \bibinfo
  {author} {\bibfnamefont {T.~J.~A.}\ \bibnamefont {Wolf}}, \bibinfo {author}
  {\bibfnamefont {Z.}~\bibnamefont {Li}}, \bibinfo {author} {\bibfnamefont
  {J.~P.~F.}\ \bibnamefont {Nunes}}, \bibinfo {author} {\bibfnamefont
  {R.}~\bibnamefont {Coffee}}, \bibinfo {author} {\bibfnamefont {J.~P.}\
  \bibnamefont {Cryan}}, \bibinfo {author} {\bibfnamefont {M.}~\bibnamefont
  {Gühr}}, \bibinfo {author} {\bibfnamefont {K.}~\bibnamefont {Hegazy}},
  \bibinfo {author} {\bibfnamefont {T.~F.}\ \bibnamefont {Heinz}}, \bibinfo
  {author} {\bibfnamefont {K.}~\bibnamefont {Jobe}}, \bibinfo {author}
  {\bibfnamefont {R.}~\bibnamefont {Li}}, \bibinfo {author} {\bibfnamefont
  {X.}~\bibnamefont {Shen}}, \bibinfo {author} {\bibfnamefont {T.}~\bibnamefont
  {Veccione}}, \bibinfo {author} {\bibfnamefont {S.}~\bibnamefont
  {Weathersby}}, \bibinfo {author} {\bibfnamefont {K.~J.}\ \bibnamefont
  {Wilkin}}, \bibinfo {author} {\bibfnamefont {C.}~\bibnamefont {Yoneda}},
  \bibinfo {author} {\bibfnamefont {Q.}~\bibnamefont {Zheng}}, \bibinfo
  {author} {\bibfnamefont {T.~J.}\ \bibnamefont {Martinez}}, \bibinfo {author}
  {\bibfnamefont {M.}~\bibnamefont {Centurion}},\ and\ \bibinfo {author}
  {\bibfnamefont {X.}~\bibnamefont {Wang}},\ }\bibfield  {title} {\enquote
  {\bibinfo {title} {Imaging \ce{CF3I} conical intersection and
  photodissociation dynamics with ultrafast electron diffraction},}\
  }\href@noop {} {\bibfield  {journal} {\bibinfo  {journal} {Science}\ }\textbf
  {\bibinfo {volume} {361}},\ \bibinfo {pages} {64--67} (\bibinfo {year}
  {2018})}\BibitemShut {NoStop}%
\bibitem [{\citenamefont {Pigliucci}\ \emph {et~al.}(2007)\citenamefont
  {Pigliucci}, \citenamefont {Duvanel}, \citenamefont {Daku},\ and\
  \citenamefont {Vauthey}}]{Pigliucci2007}%
  \BibitemOpen
  \bibfield  {author} {\bibinfo {author} {\bibfnamefont {A.}~\bibnamefont
  {Pigliucci}}, \bibinfo {author} {\bibfnamefont {G.}~\bibnamefont {Duvanel}},
  \bibinfo {author} {\bibfnamefont {L.~M.~L.}\ \bibnamefont {Daku}},\ and\
  \bibinfo {author} {\bibfnamefont {E.}~\bibnamefont {Vauthey}},\ }\bibfield
  {title} {\enquote {\bibinfo {title} {Investigation of the influence of
  solute-solvent interactions on the vibrational energy relaxation dynamics of
  large molecules in liquids},}\ }\href@noop {} {\bibfield  {journal} {\bibinfo
   {journal} {J. Phys. Chem. A}\ }\textbf {\bibinfo {volume} {111}},\ \bibinfo
  {pages} {6135--6145} (\bibinfo {year} {2007})}\BibitemShut {NoStop}%
\bibitem [{\citenamefont {Underwood}\ and\ \citenamefont
  {Blank}(2003)}]{Underwood2003}%
  \BibitemOpen
  \bibfield  {author} {\bibinfo {author} {\bibfnamefont {D.~F.}\ \bibnamefont
  {Underwood}}\ and\ \bibinfo {author} {\bibfnamefont {D.~A.}\ \bibnamefont
  {Blank}},\ }\bibfield  {title} {\enquote {\bibinfo {title} {Ultrafast
  solvation dynamics: A view from the solvent's perspective using a novel
  resonant-pump, nonresonant-probe technique},}\ }\href@noop {} {\bibfield
  {journal} {\bibinfo  {journal} {J. Phys. Chem. A}\ }\textbf {\bibinfo
  {volume} {107}},\ \bibinfo {pages} {956--961} (\bibinfo {year}
  {2003})}\BibitemShut {NoStop}%
\bibitem [{\citenamefont {Kwak}\ \emph {et~al.}(2006)\citenamefont {Kwak},
  \citenamefont {Zheng}, \citenamefont {Cang},\ and\ \citenamefont
  {Fayer}}]{Kwak2006}%
  \BibitemOpen
  \bibfield  {author} {\bibinfo {author} {\bibfnamefont {K.}~\bibnamefont
  {Kwak}}, \bibinfo {author} {\bibfnamefont {J.}~\bibnamefont {Zheng}},
  \bibinfo {author} {\bibfnamefont {H.}~\bibnamefont {Cang}},\ and\ \bibinfo
  {author} {\bibfnamefont {M.~D.}\ \bibnamefont {Fayer}},\ }\bibfield  {title}
  {\enquote {\bibinfo {title} {Ultrafast two-dimensional infrared vibrational
  echo chemical exchange experiments and theory},}\ }\href@noop {} {\bibfield
  {journal} {\bibinfo  {journal} {J. Phys. Chem. B}\ }\textbf {\bibinfo
  {volume} {110}},\ \bibinfo {pages} {19998--20013} (\bibinfo {year}
  {2006})}\BibitemShut {NoStop}%
\bibitem [{\citenamefont {{Charles Williamson}}\ and\ \citenamefont
  {Zewail}(1993)}]{Williamson1993}%
  \BibitemOpen
  \bibfield  {author} {\bibinfo {author} {\bibfnamefont {J.}~\bibnamefont
  {{Charles Williamson}}}\ and\ \bibinfo {author} {\bibfnamefont {A.~H.}\
  \bibnamefont {Zewail}},\ }\bibfield  {title} {\enquote {\bibinfo {title}
  {Ultrafast electron diffraction. velocity mismatch and temporal resolution in
  crossed-beam experiments},}\ }\href@noop {} {\bibfield  {journal} {\bibinfo
  {journal} {Chemical Physics Letters}\ }\textbf {\bibinfo {volume} {209}},\
  \bibinfo {pages} {10--16} (\bibinfo {year} {1993})}\BibitemShut {NoStop}%
\bibitem [{\citenamefont {Bürgi}(1989)}]{Burgi_1989}%
  \BibitemOpen
  \bibfield  {author} {\bibinfo {author} {\bibfnamefont {H.~B.}\ \bibnamefont
  {Bürgi}},\ }\bibfield  {title} {\enquote {\bibinfo {title} {Stereochemical
  applications of gas-phase electron diffraction},}\ }\href@noop {} {\bibfield
  {journal} {\bibinfo  {journal} {Angew. Chem. Int. Ed. Engl.}\ }\textbf
  {\bibinfo {volume} {28}},\ \bibinfo {pages} {1534--1535} (\bibinfo {year}
  {1989})}\BibitemShut {NoStop}%
\bibitem [{\citenamefont {Ischenko}, \citenamefont {Ewbank},\ and\
  \citenamefont {Schaefer}(1994)}]{Ischenko1994}%
  \BibitemOpen
  \bibfield  {author} {\bibinfo {author} {\bibfnamefont {A.~A.}\ \bibnamefont
  {Ischenko}}, \bibinfo {author} {\bibfnamefont {J.~D.}\ \bibnamefont
  {Ewbank}},\ and\ \bibinfo {author} {\bibfnamefont {L.}~\bibnamefont
  {Schaefer}},\ }\bibfield  {title} {\enquote {\bibinfo {title} {Direct
  evaluation of equilibrium molecular geometries using real-time gas electron
  diffraction},}\ }\href@noop {} {\bibfield  {journal} {\bibinfo  {journal} {J.
  Phys. Chem.}\ }\textbf {\bibinfo {volume} {98}},\ \bibinfo {pages}
  {4287--4300} (\bibinfo {year} {1994})}\BibitemShut {NoStop}%
\bibitem [{\citenamefont {Maggard}\ \emph {et~al.}(1995)\citenamefont
  {Maggard}, \citenamefont {Lobastov}, \citenamefont {Schaefer}, \citenamefont
  {Ewbank},\ and\ \citenamefont {Ischenko}}]{Maggard1995}%
  \BibitemOpen
  \bibfield  {author} {\bibinfo {author} {\bibfnamefont {P.}~\bibnamefont
  {Maggard}}, \bibinfo {author} {\bibfnamefont {V.~A.}\ \bibnamefont
  {Lobastov}}, \bibinfo {author} {\bibfnamefont {L.}~\bibnamefont {Schaefer}},
  \bibinfo {author} {\bibfnamefont {J.~D.}\ \bibnamefont {Ewbank}},\ and\
  \bibinfo {author} {\bibfnamefont {A.~A.}\ \bibnamefont {Ischenko}},\
  }\bibfield  {title} {\enquote {\bibinfo {title} {Direct evaluation of
  equilibrium molecular geometries using real-time gas electron diffraction. 2.
  selenium hexafluoride},}\ }\href@noop {} {\bibfield  {journal} {\bibinfo
  {journal} {J. Phys. Chem.}\ }\textbf {\bibinfo {volume} {99}},\ \bibinfo
  {pages} {13115--13117} (\bibinfo {year} {1995})}\BibitemShut {NoStop}%
\bibitem [{\citenamefont {Hensley}, \citenamefont {Yang},\ and\ \citenamefont
  {Centurion}(2012)}]{Hensley2012}%
  \BibitemOpen
  \bibfield  {author} {\bibinfo {author} {\bibfnamefont {C.~J.}\ \bibnamefont
  {Hensley}}, \bibinfo {author} {\bibfnamefont {J.}~\bibnamefont {Yang}},\ and\
  \bibinfo {author} {\bibfnamefont {M.}~\bibnamefont {Centurion}},\ }\bibfield
  {title} {\enquote {\bibinfo {title} {Imaging of isolated molecules with
  ultrafast electron pulses},}\ }\href@noop {} {\bibfield  {journal} {\bibinfo
  {journal} {Phys. Rev. Lett.}\ }\textbf {\bibinfo {volume} {109}},\ \bibinfo
  {pages} {133202} (\bibinfo {year} {2012})}\BibitemShut {NoStop}%
\bibitem [{\citenamefont {Vecchione}\ \emph {et~al.}(2017)\citenamefont
  {Vecchione}, \citenamefont {Denes}, \citenamefont {Jobe}, \citenamefont
  {Johnson}, \citenamefont {Joseph}, \citenamefont {Li}, \citenamefont
  {Perazzo}, \citenamefont {Shen}, \citenamefont {Wang}, \citenamefont
  {Weathersby}, \citenamefont {Yang},\ and\ \citenamefont
  {Zhang}}]{Vecchione2017}%
  \BibitemOpen
  \bibfield  {author} {\bibinfo {author} {\bibfnamefont {T.}~\bibnamefont
  {Vecchione}}, \bibinfo {author} {\bibfnamefont {P.}~\bibnamefont {Denes}},
  \bibinfo {author} {\bibfnamefont {R.~K.}\ \bibnamefont {Jobe}}, \bibinfo
  {author} {\bibfnamefont {I.~J.}\ \bibnamefont {Johnson}}, \bibinfo {author}
  {\bibfnamefont {J.~M.}\ \bibnamefont {Joseph}}, \bibinfo {author}
  {\bibfnamefont {R.~K.}\ \bibnamefont {Li}}, \bibinfo {author} {\bibfnamefont
  {A.}~\bibnamefont {Perazzo}}, \bibinfo {author} {\bibfnamefont
  {X.}~\bibnamefont {Shen}}, \bibinfo {author} {\bibfnamefont {X.~J.}\
  \bibnamefont {Wang}}, \bibinfo {author} {\bibfnamefont {S.~P.}\ \bibnamefont
  {Weathersby}}, \bibinfo {author} {\bibfnamefont {J.}~\bibnamefont {Yang}},\
  and\ \bibinfo {author} {\bibfnamefont {D.}~\bibnamefont {Zhang}},\ }\bibfield
   {title} {\enquote {\bibinfo {title} {A direct electron detector for
  time-resolved {MeV} electron microscopy},}\ }\href@noop {} {\bibfield
  {journal} {\bibinfo  {journal} {Review of Scientific Instruments}\ }\textbf
  {\bibinfo {volume} {88}},\ \bibinfo {pages} {033702} (\bibinfo {year}
  {2017})}\BibitemShut {NoStop}%
\bibitem [{\citenamefont {Centurion}, \citenamefont {Wolf},\ and\ \citenamefont
  {Yang}(2022)}]{Centurion2022}%
  \BibitemOpen
  \bibfield  {author} {\bibinfo {author} {\bibfnamefont {M.}~\bibnamefont
  {Centurion}}, \bibinfo {author} {\bibfnamefont {T.~J.}\ \bibnamefont
  {Wolf}},\ and\ \bibinfo {author} {\bibfnamefont {J.}~\bibnamefont {Yang}},\
  }\bibfield  {title} {\enquote {\bibinfo {title} {Ultrafast imaging of
  molecules with electron diffraction},}\ }\href@noop {} {\bibfield  {journal}
  {\bibinfo  {journal} {Annual Review of Physical Chemistry}\ }\textbf
  {\bibinfo {volume} {73}},\ \bibinfo {pages} {21--42} (\bibinfo {year}
  {2022})}\BibitemShut {NoStop}%
\bibitem [{\citenamefont {Yang}\ \emph {et~al.}(2016)\citenamefont {Yang},
  \citenamefont {Guehr}, \citenamefont {Vecchione}, \citenamefont {Robinson},
  \citenamefont {Li}, \citenamefont {Hartmann}, \citenamefont {Shen},
  \citenamefont {Coffee}, \citenamefont {Corbett}, \citenamefont {Fry},
  \citenamefont {Gaffney}, \citenamefont {Gorkhover}, \citenamefont {Hast},
  \citenamefont {Jobe}, \citenamefont {Makasyuk}, \citenamefont {Reid},
  \citenamefont {Robinson}, \citenamefont {Vetter}, \citenamefont {Wang},
  \citenamefont {Weathersby}, \citenamefont {Yoneda}, \citenamefont
  {Centurion},\ and\ \citenamefont {Wang}}]{Yang2016}%
  \BibitemOpen
  \bibfield  {author} {\bibinfo {author} {\bibfnamefont {J.}~\bibnamefont
  {Yang}}, \bibinfo {author} {\bibfnamefont {M.}~\bibnamefont {Guehr}},
  \bibinfo {author} {\bibfnamefont {T.}~\bibnamefont {Vecchione}}, \bibinfo
  {author} {\bibfnamefont {M.~S.}\ \bibnamefont {Robinson}}, \bibinfo {author}
  {\bibfnamefont {R.}~\bibnamefont {Li}}, \bibinfo {author} {\bibfnamefont
  {N.}~\bibnamefont {Hartmann}}, \bibinfo {author} {\bibfnamefont
  {X.}~\bibnamefont {Shen}}, \bibinfo {author} {\bibfnamefont {R.}~\bibnamefont
  {Coffee}}, \bibinfo {author} {\bibfnamefont {J.}~\bibnamefont {Corbett}},
  \bibinfo {author} {\bibfnamefont {A.}~\bibnamefont {Fry}}, \bibinfo {author}
  {\bibfnamefont {K.}~\bibnamefont {Gaffney}}, \bibinfo {author} {\bibfnamefont
  {T.}~\bibnamefont {Gorkhover}}, \bibinfo {author} {\bibfnamefont
  {C.}~\bibnamefont {Hast}}, \bibinfo {author} {\bibfnamefont {K.}~\bibnamefont
  {Jobe}}, \bibinfo {author} {\bibfnamefont {I.}~\bibnamefont {Makasyuk}},
  \bibinfo {author} {\bibfnamefont {A.}~\bibnamefont {Reid}}, \bibinfo {author}
  {\bibfnamefont {J.}~\bibnamefont {Robinson}}, \bibinfo {author}
  {\bibfnamefont {S.}~\bibnamefont {Vetter}}, \bibinfo {author} {\bibfnamefont
  {F.}~\bibnamefont {Wang}}, \bibinfo {author} {\bibfnamefont {S.}~\bibnamefont
  {Weathersby}}, \bibinfo {author} {\bibfnamefont {C.}~\bibnamefont {Yoneda}},
  \bibinfo {author} {\bibfnamefont {M.}~\bibnamefont {Centurion}},\ and\
  \bibinfo {author} {\bibfnamefont {X.}~\bibnamefont {Wang}},\ }\bibfield
  {title} {\enquote {\bibinfo {title} {Diffractive imaging of a rotational
  wavepacket in nitrogen molecules with femtosecond megaelectronvolt electron
  pulses},}\ }\href@noop {} {\bibfield  {journal} {\bibinfo  {journal} {Nature
  Communications}\ }\textbf {\bibinfo {volume} {7}} (\bibinfo {year}
  {2016})}\BibitemShut {NoStop}%
\bibitem [{\citenamefont {Fitzpatrick}, \citenamefont {Vanacore},\ and\
  \citenamefont {Zewail}(2015)}]{Fitzpatrick_2015}%
  \BibitemOpen
  \bibfield  {author} {\bibinfo {author} {\bibfnamefont {A.~W.~P.}\
  \bibnamefont {Fitzpatrick}}, \bibinfo {author} {\bibfnamefont {G.~M.}\
  \bibnamefont {Vanacore}},\ and\ \bibinfo {author} {\bibfnamefont {A.~H.}\
  \bibnamefont {Zewail}},\ }\bibfield  {title} {\enquote {\bibinfo {title}
  {Nanomechanics and intermolecular forces of amyloid revealed by
  four-dimensional electron microscopy},}\ }\href@noop {} {\bibfield  {journal}
  {\bibinfo  {journal} {Proceedings of the National Academy of Sciences}\
  }\textbf {\bibinfo {volume} {112}},\ \bibinfo {pages} {3380--3385} (\bibinfo
  {year} {2015})}\BibitemShut {NoStop}%
\bibitem [{\citenamefont {Fitzpatrick}\ \emph {et~al.}(2013)\citenamefont
  {Fitzpatrick}, \citenamefont {Lorenz}, \citenamefont {Vanacore},\ and\
  \citenamefont {Zewail}}]{Fitzpatrick_2013}%
  \BibitemOpen
  \bibfield  {author} {\bibinfo {author} {\bibfnamefont {A.~W.~P.}\
  \bibnamefont {Fitzpatrick}}, \bibinfo {author} {\bibfnamefont {U.~J.}\
  \bibnamefont {Lorenz}}, \bibinfo {author} {\bibfnamefont {G.~M.}\
  \bibnamefont {Vanacore}},\ and\ \bibinfo {author} {\bibfnamefont {A.~H.}\
  \bibnamefont {Zewail}},\ }\bibfield  {title} {\enquote {\bibinfo {title} {4d
  cryo-electron microscopy of proteins},}\ }\href@noop {} {\bibfield  {journal}
  {\bibinfo  {journal} {Journal of the American Chemical Society}\ }\textbf
  {\bibinfo {volume} {135}},\ \bibinfo {pages} {19123--19126} (\bibinfo {year}
  {2013})}\BibitemShut {NoStop}%
\bibitem [{\citenamefont {Flannigan}, \citenamefont {Barwick},\ and\
  \citenamefont {Zewail}(2010)}]{Flannigan_2010}%
  \BibitemOpen
  \bibfield  {author} {\bibinfo {author} {\bibfnamefont {D.~J.}\ \bibnamefont
  {Flannigan}}, \bibinfo {author} {\bibfnamefont {B.}~\bibnamefont {Barwick}},\
  and\ \bibinfo {author} {\bibfnamefont {A.~H.}\ \bibnamefont {Zewail}},\
  }\bibfield  {title} {\enquote {\bibinfo {title} {Biological imaging with 4d
  ultrafast electron microscopy},}\ }\href@noop {} {\bibfield  {journal}
  {\bibinfo  {journal} {Proceedings of the National Academy of Sciences}\
  }\textbf {\bibinfo {volume} {107}},\ \bibinfo {pages} {9933--9937} (\bibinfo
  {year} {2010})}\BibitemShut {NoStop}%
\bibitem [{\citenamefont {Henderson}(1995)}]{Henderson_1995}%
  \BibitemOpen
  \bibfield  {author} {\bibinfo {author} {\bibfnamefont {R.}~\bibnamefont
  {Henderson}},\ }\bibfield  {title} {\enquote {\bibinfo {title} {The potential
  and limitations of neutrons, electrons and x-rays for atomic resolution
  microscopy of unstained biological molecules},}\ }\href@noop {} {\bibfield
  {journal} {\bibinfo  {journal} {Quarterly Reviews of Biophysics}\ }\textbf
  {\bibinfo {volume} {28}},\ \bibinfo {pages} {171--193} (\bibinfo {year}
  {1995})}\BibitemShut {NoStop}%
\bibitem [{\citenamefont {Weathersby}\ \emph {et~al.}(2015)\citenamefont
  {Weathersby}, \citenamefont {Brown}, \citenamefont {Centurion}, \citenamefont
  {Chase}, \citenamefont {Coffee}, \citenamefont {Corbett}, \citenamefont
  {Eichner}, \citenamefont {Frisch}, \citenamefont {Fry}, \citenamefont
  {Gühr}, \citenamefont {Hartmann}, \citenamefont {Hast}, \citenamefont
  {Hettel}, \citenamefont {Jobe}, \citenamefont {Jongewaard}, \citenamefont
  {Lewandowski}, \citenamefont {Li}, \citenamefont {Lindenberg}, \citenamefont
  {Makasyuk}, \citenamefont {May}, \citenamefont {McCormick}, \citenamefont
  {Nguyen}, \citenamefont {Reid}, \citenamefont {Shen}, \citenamefont
  {Sokolowski-Tinten}, \citenamefont {Vecchione}, \citenamefont {Vetter},
  \citenamefont {Wu}, \citenamefont {Yang}, \citenamefont {Dürr},\ and\
  \citenamefont {Wang}}]{Weathersby2015}%
  \BibitemOpen
  \bibfield  {author} {\bibinfo {author} {\bibfnamefont {S.~P.}\ \bibnamefont
  {Weathersby}}, \bibinfo {author} {\bibfnamefont {G.}~\bibnamefont {Brown}},
  \bibinfo {author} {\bibfnamefont {M.}~\bibnamefont {Centurion}}, \bibinfo
  {author} {\bibfnamefont {T.~F.}\ \bibnamefont {Chase}}, \bibinfo {author}
  {\bibfnamefont {R.}~\bibnamefont {Coffee}}, \bibinfo {author} {\bibfnamefont
  {J.}~\bibnamefont {Corbett}}, \bibinfo {author} {\bibfnamefont {J.~P.}\
  \bibnamefont {Eichner}}, \bibinfo {author} {\bibfnamefont {J.~C.}\
  \bibnamefont {Frisch}}, \bibinfo {author} {\bibfnamefont {A.~R.}\
  \bibnamefont {Fry}}, \bibinfo {author} {\bibfnamefont {M.}~\bibnamefont
  {Gühr}}, \bibinfo {author} {\bibfnamefont {N.}~\bibnamefont {Hartmann}},
  \bibinfo {author} {\bibfnamefont {C.}~\bibnamefont {Hast}}, \bibinfo {author}
  {\bibfnamefont {R.}~\bibnamefont {Hettel}}, \bibinfo {author} {\bibfnamefont
  {R.~K.}\ \bibnamefont {Jobe}}, \bibinfo {author} {\bibfnamefont {E.~N.}\
  \bibnamefont {Jongewaard}}, \bibinfo {author} {\bibfnamefont {J.~R.}\
  \bibnamefont {Lewandowski}}, \bibinfo {author} {\bibfnamefont {R.~K.}\
  \bibnamefont {Li}}, \bibinfo {author} {\bibfnamefont {A.~M.}\ \bibnamefont
  {Lindenberg}}, \bibinfo {author} {\bibfnamefont {I.}~\bibnamefont
  {Makasyuk}}, \bibinfo {author} {\bibfnamefont {J.~E.}\ \bibnamefont {May}},
  \bibinfo {author} {\bibfnamefont {D.}~\bibnamefont {McCormick}}, \bibinfo
  {author} {\bibfnamefont {M.~N.}\ \bibnamefont {Nguyen}}, \bibinfo {author}
  {\bibfnamefont {A.~H.}\ \bibnamefont {Reid}}, \bibinfo {author}
  {\bibfnamefont {X.}~\bibnamefont {Shen}}, \bibinfo {author} {\bibfnamefont
  {K.}~\bibnamefont {Sokolowski-Tinten}}, \bibinfo {author} {\bibfnamefont
  {T.}~\bibnamefont {Vecchione}}, \bibinfo {author} {\bibfnamefont {S.~L.}\
  \bibnamefont {Vetter}}, \bibinfo {author} {\bibfnamefont {J.}~\bibnamefont
  {Wu}}, \bibinfo {author} {\bibfnamefont {J.}~\bibnamefont {Yang}}, \bibinfo
  {author} {\bibfnamefont {H.~A.}\ \bibnamefont {Dürr}},\ and\ \bibinfo
  {author} {\bibfnamefont {X.~J.}\ \bibnamefont {Wang}},\ }\bibfield  {title}
  {\enquote {\bibinfo {title} {Mega-electron-volt ultrafast electron
  diffraction at {SLAC} {National} {Accelerator} {Laboratory}},}\ }\href@noop
  {} {\bibfield  {journal} {\bibinfo  {journal} {Review of Scientific
  Instruments}\ }\textbf {\bibinfo {volume} {86}},\ \bibinfo {pages} {073702}
  (\bibinfo {year} {2015})}\BibitemShut {NoStop}%
\bibitem [{\citenamefont {van Oudheusden}\ \emph {et~al.}(2010)\citenamefont
  {van Oudheusden}, \citenamefont {Pasmans}, \citenamefont {van~der Geer},
  \citenamefont {de~Loos}, \citenamefont {van~der Wiel},\ and\ \citenamefont
  {Luiten}}]{Luiten2010}%
  \BibitemOpen
  \bibfield  {author} {\bibinfo {author} {\bibfnamefont {T.}~\bibnamefont {van
  Oudheusden}}, \bibinfo {author} {\bibfnamefont {P.~L. E.~M.}\ \bibnamefont
  {Pasmans}}, \bibinfo {author} {\bibfnamefont {S.~B.}\ \bibnamefont {van~der
  Geer}}, \bibinfo {author} {\bibfnamefont {M.~J.}\ \bibnamefont {de~Loos}},
  \bibinfo {author} {\bibfnamefont {M.~J.}\ \bibnamefont {van~der Wiel}},\ and\
  \bibinfo {author} {\bibfnamefont {O.~J.}\ \bibnamefont {Luiten}},\ }\bibfield
   {title} {\enquote {\bibinfo {title} {Compression of subrelativistic
  space-charge-dominated electron bunches for single-shot femtosecond electron
  diffraction},}\ }\href@noop {} {\bibfield  {journal} {\bibinfo  {journal}
  {Phys. Rev. Lett.}\ }\textbf {\bibinfo {volume} {105}},\ \bibinfo {pages}
  {264801} (\bibinfo {year} {2010})}\BibitemShut {NoStop}%
\bibitem [{\citenamefont {Mo}\ \emph {et~al.}(2016{\natexlab{b}})\citenamefont
  {Mo}, \citenamefont {Shen}, \citenamefont {Chen}, \citenamefont {Li},
  \citenamefont {Dunning}, \citenamefont {Sokolowski-Tinten}, \citenamefont
  {Zheng}, \citenamefont {Weathersby}, \citenamefont {Reid}, \citenamefont
  {Coffee}, \citenamefont {Makasyuk}, \citenamefont {Edstrom}, \citenamefont
  {McCormick}, \citenamefont {Jobe}, \citenamefont {Hast}, \citenamefont
  {Glenzer},\ and\ \citenamefont {Wang}}]{Mo_2016}%
  \BibitemOpen
  \bibfield  {author} {\bibinfo {author} {\bibfnamefont {M.~Z.}\ \bibnamefont
  {Mo}}, \bibinfo {author} {\bibfnamefont {X.}~\bibnamefont {Shen}}, \bibinfo
  {author} {\bibfnamefont {Z.}~\bibnamefont {Chen}}, \bibinfo {author}
  {\bibfnamefont {R.~K.}\ \bibnamefont {Li}}, \bibinfo {author} {\bibfnamefont
  {M.}~\bibnamefont {Dunning}}, \bibinfo {author} {\bibfnamefont
  {K.}~\bibnamefont {Sokolowski-Tinten}}, \bibinfo {author} {\bibfnamefont
  {Q.}~\bibnamefont {Zheng}}, \bibinfo {author} {\bibfnamefont {S.~P.}\
  \bibnamefont {Weathersby}}, \bibinfo {author} {\bibfnamefont {A.~H.}\
  \bibnamefont {Reid}}, \bibinfo {author} {\bibfnamefont {R.}~\bibnamefont
  {Coffee}}, \bibinfo {author} {\bibfnamefont {I.}~\bibnamefont {Makasyuk}},
  \bibinfo {author} {\bibfnamefont {S.}~\bibnamefont {Edstrom}}, \bibinfo
  {author} {\bibfnamefont {D.}~\bibnamefont {McCormick}}, \bibinfo {author}
  {\bibfnamefont {K.}~\bibnamefont {Jobe}}, \bibinfo {author} {\bibfnamefont
  {C.}~\bibnamefont {Hast}}, \bibinfo {author} {\bibfnamefont {S.~H.}\
  \bibnamefont {Glenzer}},\ and\ \bibinfo {author} {\bibfnamefont
  {X.}~\bibnamefont {Wang}},\ }\bibfield  {title} {\enquote {\bibinfo {title}
  {Single-shot mega-electronvolt ultrafast electron diffraction for structure
  dynamic studies of warm dense matter},}\ }\href@noop {} {\bibfield  {journal}
  {\bibinfo  {journal} {Review of Scientific Instruments}\ }\textbf {\bibinfo
  {volume} {87}},\ \bibinfo {pages} {11D810} (\bibinfo {year}
  {2016}{\natexlab{b}})}\BibitemShut {NoStop}%
\bibitem [{\citenamefont {Qi}\ \emph {et~al.}(2020)\citenamefont {Qi},
  \citenamefont {Ma}, \citenamefont {Zhao}, \citenamefont {Cheng},
  \citenamefont {Jiang}, \citenamefont {Lu}, \citenamefont {Jiang},
  \citenamefont {Qian}, \citenamefont {Wang}, \citenamefont {Zhang},
  \citenamefont {Zhu}, \citenamefont {Zou}, \citenamefont {Wan}, \citenamefont
  {Xiang},\ and\ \citenamefont {Zhang}}]{PhysRevLett.124.134803}%
  \BibitemOpen
  \bibfield  {author} {\bibinfo {author} {\bibfnamefont {F.}~\bibnamefont
  {Qi}}, \bibinfo {author} {\bibfnamefont {Z.}~\bibnamefont {Ma}}, \bibinfo
  {author} {\bibfnamefont {L.}~\bibnamefont {Zhao}}, \bibinfo {author}
  {\bibfnamefont {Y.}~\bibnamefont {Cheng}}, \bibinfo {author} {\bibfnamefont
  {W.}~\bibnamefont {Jiang}}, \bibinfo {author} {\bibfnamefont
  {C.}~\bibnamefont {Lu}}, \bibinfo {author} {\bibfnamefont {T.}~\bibnamefont
  {Jiang}}, \bibinfo {author} {\bibfnamefont {D.}~\bibnamefont {Qian}},
  \bibinfo {author} {\bibfnamefont {Z.}~\bibnamefont {Wang}}, \bibinfo {author}
  {\bibfnamefont {W.}~\bibnamefont {Zhang}}, \bibinfo {author} {\bibfnamefont
  {P.}~\bibnamefont {Zhu}}, \bibinfo {author} {\bibfnamefont {X.}~\bibnamefont
  {Zou}}, \bibinfo {author} {\bibfnamefont {W.}~\bibnamefont {Wan}}, \bibinfo
  {author} {\bibfnamefont {D.}~\bibnamefont {Xiang}},\ and\ \bibinfo {author}
  {\bibfnamefont {J.}~\bibnamefont {Zhang}},\ }\bibfield  {title} {\enquote
  {\bibinfo {title} {Breaking 50 femtosecond resolution barrier in {MeV}
  ultrafast electron diffraction with a double bend achromat compressor},}\
  }\href@noop {} {\bibfield  {journal} {\bibinfo  {journal} {Phys. Rev. Lett.}\
  }\textbf {\bibinfo {volume} {124}},\ \bibinfo {pages} {134803} (\bibinfo
  {year} {2020})}\BibitemShut {NoStop}%
\bibitem [{\citenamefont {Kim}\ \emph {et~al.}(2020)\citenamefont {Kim},
  \citenamefont {Vinokurov}, \citenamefont {Baek}, \citenamefont {Oang},
  \citenamefont {Kim}, \citenamefont {Kim}, \citenamefont {Jang}, \citenamefont
  {Lee}, \citenamefont {Park}, \citenamefont {Park} \emph
  {et~al.}}]{kim2020towards}%
  \BibitemOpen
  \bibfield  {author} {\bibinfo {author} {\bibfnamefont {H.~W.}\ \bibnamefont
  {Kim}}, \bibinfo {author} {\bibfnamefont {N.~A.}\ \bibnamefont {Vinokurov}},
  \bibinfo {author} {\bibfnamefont {I.~H.}\ \bibnamefont {Baek}}, \bibinfo
  {author} {\bibfnamefont {K.~Y.}\ \bibnamefont {Oang}}, \bibinfo {author}
  {\bibfnamefont {M.~H.}\ \bibnamefont {Kim}}, \bibinfo {author} {\bibfnamefont
  {Y.~C.}\ \bibnamefont {Kim}}, \bibinfo {author} {\bibfnamefont {K.-H.}\
  \bibnamefont {Jang}}, \bibinfo {author} {\bibfnamefont {K.}~\bibnamefont
  {Lee}}, \bibinfo {author} {\bibfnamefont {S.~H.}\ \bibnamefont {Park}},
  \bibinfo {author} {\bibfnamefont {S.}~\bibnamefont {Park}}, \emph {et~al.},\
  }\bibfield  {title} {\enquote {\bibinfo {title} {Towards jitter-free
  ultrafast electron diffraction technology},}\ }\href@noop {} {\bibfield
  {journal} {\bibinfo  {journal} {Nature photonics}\ }\textbf {\bibinfo
  {volume} {14}},\ \bibinfo {pages} {245--249} (\bibinfo {year}
  {2020})}\BibitemShut {NoStop}%
\end{thebibliography}%
\end{document}